\newcommand{\bfu}{{\bf u}}

\newcommand{\bfX}{{\bf X}}
\newcommand{\bfU}{{\bf U}}
\newcommand{\bfe}{{\bf e}}

\documentclass{ar-1col}
\usepackage{mathtools}
\usepackage{newtxtext}
\usepackage{amsmath}
\usepackage{amssymb}
\usepackage{bm}
\usepackage{subfigure}
\usepackage{graphicx}
\usepackage{placeins}
\usepackage{color}
\usepackage[sort&compress]{natbib}
\usepackage[textwidth=2.1cm]{todonotes}
\usepackage{soul}
\usepackage{amsmath,amsfonts,mathtools}
\usepackage{bm}

\DeclareSymbolFont{matha}{OML}{txmi}{m}{it}
\DeclareMathSymbol{\varv}{\mathord}{matha}{118}

\jname{Annu. Rev. Condens. Matter Phys.}
\jvol{~ ~~~ ~  ~ 11~529-559} 
\jyear{2020}
\doi{https://doi.org/10.1146/annurev-conmatphys-031119-050637}

\usepackage[left=3.7cm, right=3.7cm, top=3cm]{geometry}
\begin{document}

\markboth{Mathai et al. Bubbles and Buoyant Particles in Turbulence}{}

\title{\normalfont{~\\ ~\\ ~\\~\\Bubbly and Buoyant Particle-Laden Turbulent Flows}}

\author{{\normalfont Varghese Mathai$^{1}$, Detlef Lohse$^{2,3}$, Chao Sun$^{4}$} 
	\affil{$^1$School of Engineering, Brown University, Providence, RI 02912, USA}
	\affil{$^2$Physics of Fluids Group, Max-Planck-Center Twente for Complex Fluid Dynamics, Mesa+ Institute, and J. M. Burgers Centre for Fluid Dynamics, Department of Science and Technology, University of Twente, 7500 AE, Enschede, The Netherlands}
	\affil{$^3$Max Planck Institute for Dynamics and Self-Organization, Am Fassberg 17, 37077 G\"ottingen, Germany}
	\affil{$^4$Center for Combustion Energy, Key Laboratory for Thermal Science and Power Engineering of Ministry of Education, Department of Energy and Power Engineering,~Tsinghua~University,~Beijing,~China.~Email:~chaosun@tsinghua.edu.cn}}

\begin{abstract}
\vspace{-2mm}
Fluid turbulence is commonly associated with stronger drag, greater heat transfer, and more efficient mixing than in laminar flows. In many natural and industrial settings, turbulent liquid flows contain suspensions of dispersed bubbles and light particles. Recently, much attention has been devoted to understanding the behavior and underlying physics of such flows by use of both experiments and high-resolution direct numerical simulations. This review summarizes our present understanding of various phenomenological aspects of bubbly and buoyant particle–laden turbulent flows. We begin by discussing different dynamical regimes, including those of crossing trajectories and wake-induced oscillations of rising particles, and regimes in which bubbles and particles preferentially accumulate near walls or within vortical structures. We then address how certain paradigmatic turbulent flows, such as homogeneous isotropic turbulence, channel flow, Taylor–Couette turbulence, and thermally driven turbulence, are modified by the presence of these dispersed bubbles and buoyant particles. We end with a list of summary points and future research questions.
\end{abstract}

\begin{keywords}
	\vspace{-1mm}
	Bubbles, Buoyant particles,  Lagrangian dynamics, Wake turbulence interaction, Bubble induced turbulence, Two-way coupling,
\end{keywords}
\maketitle

\FloatBarrier 

\section{INTRODUCTION AND OBJECTIVES OF REVIEW}


Turbulent multiphase flows are widely prevalent in nature and industry. Typical examples are pollutants dispersed in the atmosphere, air bubbles and plankton in the oceans, sediment laden river flows, and catalytic particles and bubble columns in process technology. In all of these
examples, the particles of the dispersed phase have a different mass density from the carrier phase. When the dispersed particle is lighter than the carrier fluid element, there can be  major consequences on the kinematics and dynamics of both phases, often triggering a multitude of flow modifications. \textcolor{black}{Many of these flow modifications can be viewed as originating from the inherent ``buoyancy'' of the bubbles and light particles, which renders them capable of adding energy over a range of scales while ascending through the turbulent flow.}
These have decisive roles in many natural phenomena we observe around us, for example in the distributions of buoyant zooplankton undergoing diel vertical migrations~\citep[e.g.][]{sengupta2017phytoplankton,calzavarini2018propelled}, in bubble induced mixing of the upper oceans \citep[e.g.][]{thorpe1987bubble}, and in engineering applications of drag reduction, heat transfer and mixing \citep[e.g.][]{ceccio2010friction,gvozdic2018experimental,almeras2015mixing}. 

The subject of  dispersed particles in turbulent flows has been studied extensively from the
perspective of small (inertial) particles. For reviews on experimental and numerical techniques for the particle laden flows, readers are referred to \cite{toschi2009lagrangian,voth2017anisotropic,crowe1996numerical,maxey2017simulation,elghobashi2019direct,prosperetti2015life,elghobashi1994predicting}. 
Over the past couple of decades, many insights has been gained on particle dynamics and flow modulations at various scales of turbulence. The vast majority of these have been on neutrally buoyant and heavy particles. As we will discuss in this review, for a variety of reasons, the heavy \textcolor{black}{and neutrally buoyant} particle laden flows are experimentally, numerically, and theoretically more amenable as compared to their bubble laden counterparts \citep{lohse2018bubble}. \cite{balachandar2010turbulent} presented the most recent review discussing important aspects of turbulent bubbly flows. Yet, these reviews do not address the subject from the viewpoint of how the ``buoyancy'' of gas bubbles and light particles can manifest in various forms in turbulent multiphase settings.  The aim of the present review is to examine the subject from this perspective, for bubbles and buoyant particles in turbulence. Therefore, all flows considered will be liquid, and we will henceforth use the term buoyant particle, generically, to refer to both gas bubbles and light particles. Vapor bubbles,  which have rich underlying physics of their own \citep{prosperetti2017vapor}, will not be discussed here.

The review is organized into two parts. In the first part, we will begin by addressing the dynamics of small bubbles and buoyant particles in a regime where particle inertia is dominant, followed by a regime where buoyancy and inertia are competing effects, leading to clustering and reduced rise velocities. We then explore the unsteady wake induced dynamics of finite sized buoyant particles and air bubbles, and later, their so-called lift and lateral migration tendencies. In the second part of the review, we discuss the main consequences of the particle dynamics on flow modulations including drag reduction and heat transfer enhancement pertaining to different turbulent flow systems. The review ends with a summary and outlook towards open issues for future research.

\subsection{Dimensionless groups}
{\color{black} To cover the vastness of issues encountered in buoyant particle laden turbulence requires an exhaustive compilation of the relevant dimensionless parameters; for a review on the general subject of bubbles see  \cite{lohse2018bubble}. In the present review, we restrict our attention to a selected subset of these.  We divide the control parameter space into particle parameters and flow parameters. For the bubble or buoyant particle, the following dimensionless groups are important: the density ratio of particle (or bubble) to liquid $\Gamma \equiv \rho_p/\rho_l$, the size ratio, $\Xi \equiv d_p/{l}$, where $d_p$  is the particle diameter and $l$ a characteristic length scale of the flow. Combining $\Gamma$ and $\Xi$, one can obtain the Stokes number St $\equiv \tau_p/\tau_l$, where $\tau_p$ is the response time of the particle and $\tau_l$ a characteristic time scale of the flow. We note that the problem is sufficiently defined once two among the three parameters [$\Gamma$, $\Xi$, St] are specified.  Further, the effect of buoyancy (or gravity) can be included by using two additional control parameters: the Froude number Fr$_\text{ug}$ and the Galileo number Ga. In the most general form\footnote{Note that many different, but equivalent, definitions of Froude number are available.} Fr$_\text{ug}$ = $U^2/(gd_p)$ gives the ratio of inertia to buoyancy, where $U$ is a relevant velocity scale and $g$ the gravitational acceleration. Ga~$\equiv~{\sqrt{g d_p^3 (\Gamma -1) }}/{\nu}$ compares buoyancy effects to viscous effects \citep{veldhuis2004motion}, where $\nu$ is the kinematic viscosity of the carrier liquid. Note that the above definition for Ga, which is alternately referred to as the Archimedes number Ar is equivalent \citep{ern2012wake}. Next, when the buoyant particle under consideration is a gas bubble, its deformability can have an influence on the particle-turbulence interactions. The E\"otv\"os number Eo = $\rho_l g d_p^2/\sigma$ (alternately known as the Bond number Bo) gives the ratio of buoyancy force to capillary force. Here $\rho_l$ is the liquid density, and $\sigma$ is the surface tension of the gas-liquid interface. Furthermore, when the buoyant particle's rotation is important, its mass moment of inertia ratio I$^* \equiv$ I$_p$/I$_l$ becomes a relevant control parameter, where I$_p$ is the particle moment of inertia and I$_l$  the moment of inertia of the displaced liquid.} For the turbulence, we will refer to the Reynolds number (with varying definitions based on the type of the turbulent flow) and the Rayleigh number Ra (for thermal turbulence). With these parameters specified, we look at a variety of responses of the particles, as well as the flow modulation they induce. 

{\color{black}
	
	Lastly, the volume fraction $\alpha$ of the dispersed phase is a crucial control parameter for the system. For particle laden systems, it is common to  specify a low volume fraction threshold for transitions to two-way and four-way coupling regimes \citep{elghobashi1994predicting}. For bubble and light particle laden turbulence, such a criterion is rarely used. As is often the case, even one buoyant bubble can significantly modify the turbulence around it, independently of the volume fraction. Thus, to point out a threshold of bubble volume fraction below which they can be considered as passive is not appropriate. In this review we restrict our discussions to situations of low and moderate volume fractions ($\alpha < 5\%$), where complex issues of coalescence and break up \citep{jha2015interaction} are not dominant.}

In addition to the above mentioned control parameters, the particle Reynolds number Re$_p \equiv v_T d_p/\nu$ and a buoyancy parameter R$_v \equiv v_T/u'$, which are estimable  output parameters, will also be discussed. Here $v_T$ is the measured mean particle rise velocity, and $u'$ is the rms velocity of liquid fluctuations of the single phase turbulence. Broadly speaking, the Lagrangian dynamics of the particles and the various flow modifications they induce will be the output quantities of importance. Our intention here is to present these under one umbrella of familiar terminology. Further details of  the above mentioned dimensionless groups will be provided in the relevant sections.

\section{PARTICLE MOTION IN A FLUID FLOW}
\label{Flow and particles}

{\color{black}The fluid motion in an incompressible multiphase flow is governed by the Navier-Stokes~(NS) equations 
	
	\begin{figure*} [!tbp]
		\centering
		\includegraphics[width=0.98\linewidth]{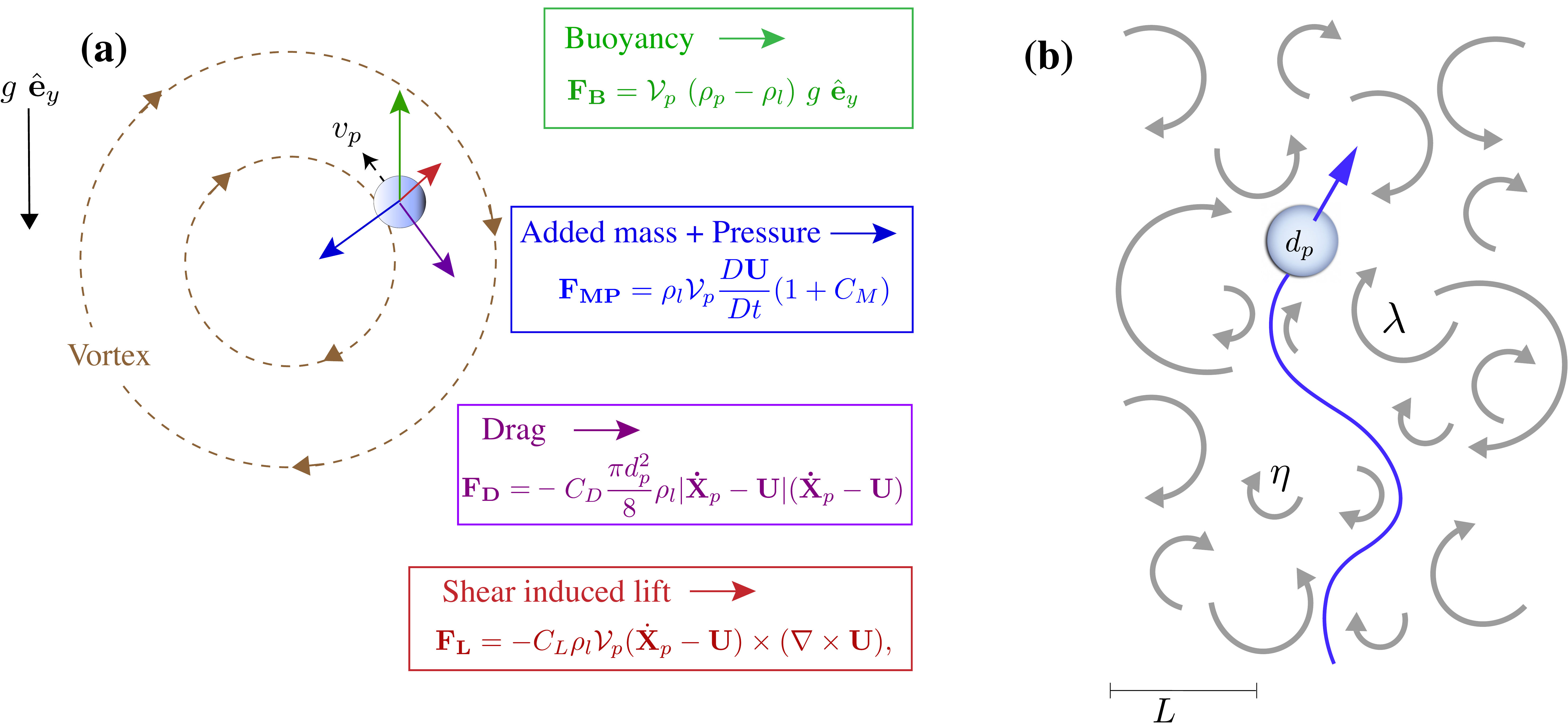}
		\caption{ Schematic representations of a buoyant spherical particle in a turbulent flow. (a) Schematic of the forces (represented as point forces) felt by a small buoyant particle (or gas bubble) rising past a vortex (a simplified representation of turbulent eddy). $v_p$ is the instantaneous particle velocity vector, and the forces due to buoyancy ${\bf F_B}$, drag ${\bf F_D}$, so-called lift ${\bf F_L}$, and the net effect of added mass and pressure ${\bf F_{MP}}$, are shown using colored arrows giving their relative directions. Here ${\bf F_{MP}}$ is obtained by rearranging the added mass and fluid force terms in eq.~(\ref{eq_dim}), and accounts for the pressure force on the buoyant particle. For details of the variables shown, we refer to Eq.~\ref{eq_fam}-\ref{eq_fd}. (b) Schematic representation of the buoyant particle rising through a turbulent flow composed of vortices (or eddies) of various length and time scales. Here $d_p$ is the particle diameter, $\eta$ is the dissipative length scale, and $L$ is the integral length scale of the flow  \citep{pop00,toschi2009lagrangian}. $\lambda$ is an intermediate length scale, commonly known as the Taylor micro-scale.}
		\label{fig:1_PIT_schematic} 
	\end{figure*}

	\begin{equation}
	\frac{\partial{{\bf U}}}{\partial{t}} + ({\bf U \cdot \nabla}){\bf U} = - \frac{\nabla p}{\rho_l} + \nu \nabla^2 {\bf U} + {\bf f_\text{R}},
	\end{equation}
} 
where ${\bf U}$ is the velocity at the location of the particle, $p$ is the pressure, $\rho_l$ is the liquid density, $\nu$ is the kinematic viscosity of the liquid, and ${\bf f_\text{R}}$ is a back reaction force per unit mass on the fluid. Based on the flow setting and the modeling approach, ${\bf f_\text{R}}$ can be either exactly computed \citep{elghobashi2019direct} or modeled \citep{mazzitelli2003effect,mazzitelli2003relevance} or ignored altogether \citep{toschi2009lagrangian}. Appropriate boundary conditions need to be applied on the boundaries of the fluid domain. For the particle, when employing fully resolved direct numerical simulations~(DNS), the boundary condition is either no-slip or free slip or a combination of the two.


We begin by considering the equation of motion of a small buoyant spherical particle advected in a fluid flow with velocity $\bfU(\bfX({t}),{t})$, and in the presence of gravity. Assuming the particle's spatial dimension is point-like, one can use a form of equations in the spirit of the celebrated Maxey-Riley equation \citep{maxey1983equation}:

\begin{equation}
\mathcal{V}_p\ \rho_p\ \ddot{\bfX}_p =   \mathcal{V}_p\ \rho_l \frac{D\bfU}{D {t}}  + \bf{F}_{M}  + \bf{F}_{B} + \bf{F}_{L} + \bf{F}_{D}, 
\label{eq_dim}
\end{equation}
where ${\bfX}_p$ is the position of the particle (or bubble), $\mathcal{V}_p =  \frac{\pi d_p^3}{6}$ its volume, and $\rho_l$ and $\rho_p$ are the liquid and particle mass densities, respectively. In obtaining the above equation, the coupling between translation and rotation of the particle has been assumed negligible. The forces contributing on the right-hand-side besides those due to the accelerated flow~(which includes the pressure gradient term) are the added mass $\bf{F_M}$, drag $\bf{F}_{D}$,  buoyancy $\bf{F}_{B}$ and a so-called (shear-induced) lift $\bf{F}_{L}$. Generally, they are modelled as


\begin{eqnarray}
{\bf F_{M}} &=&  C_M \rho_l \mathcal{V}_p \  \left( \frac{D\bfU}{D{t}} -  \ddot{\bfX}_p \right),\label{eq_fam}
\\
{\bf F_{B}} &=&  \mathcal{V}_p\ (\rho_p - \rho_l)\ g\ \hat{\bfe}_{y},\label{eq_fb}
\\
{\bf F_L} &=& - C_L \rho_l \mathcal{V}_p(\dot{\bf{X}}_p- {\bf U}) \times (\nabla \times {\bf U}),\label{eq_fl}
\\
\bf{F_D} &=& - C_D \frac{\pi d_p^2}{8} \rho_l  |{\bf \dot{X}}_p- {\bf U}| ({\bf \dot{X}}_p -{\bf U}),
\label{eq_fd}
\end{eqnarray}
where $C_M$ is the added mass coefficient, $g$ is the gravitational acceleration directed along $\hat{\bfe}_{y}$, the unit-vector in the vertical direction. The shear induced lift depends on the alignment between the vorticity vector $(\nabla \times {\bf U})$ and the relative velocity of the particle $({\bf \dot{X}}_p -{\bf U})$, and $C_L$ and $C_D$ the constants of proportionality for the lift and drag, respectively (life and drag coefficents). It is important to first appreciate the directions of these force vectors for a rising buoyant particle. A simplified picture is drawn in {\bf Figure}~\ref{fig:1_PIT_schematic}(a), through a cartoon of the buoyant particle rising along the downward side of a ``vortex''. The approximate directions of the effective forces acting on the particle are also sketched: the drag (purple arrow) is oriented to oppose the direction of the particle's velocity, the buoyancy (green arrow) is vertical, the shear-induced lift force (red arrow) acts perpendicular to the plane containing the particle velocity and the vorticity vector, and is opposed by the centrifugal force (blue arrow) that is directed towards the eye of the vortex, as the particle is lighter than the surrounding liquid. Note that the Fa\'xen forces (accounting for flow non-uniformity at the particle scale) and the Basset history force have been omitted for simplicity  \citep{auton1988force,rensen2001spiraling}. The relative importance of these terms for buoyant rising particles is still to be resolved \citep{calzavarini2012impact,calzavarini2009acceleration,homann2010finite}. 

When the condition of particle Reynolds number $Re_p \ll 1$ is met, the drag coefficient reduces to $C_D=24/\text{Re}_p$, which implies the linear drag relation $\bf{F}_{D}$ = $-3\ \pi\ \mu\ d_p\ (\dot{\bfX}_p - \bfU)$. While the basic form of this equation is founded on a unified treatment of particles, drops and bubbles, it further assumes that for Re$_p \ll 1$, the particle locally sees a Stokes flow  \citep{maxey1983equation,leal1980particle} despite the unsteadiness and turbulence in the carrier flow. The expression for the drag used above assumes a contaminated air-liquid interface for the bubble.   For a clean bubble interface, the prefactor of $\bf{F}_{D}$ is slightly modified \citep{mougin2001path}, although retaining its functional form.  The assumption of a contaminated interface is indeed reasonable in most natural and industrial flows, since the carrier liquid is almost never ultrapure. Assuming potential flow in the outer regions, we can use $C_M=1/2$. Similarly, by considering the momentum flux far from the particle, one can obtain the Auton lift $C_L = 1/2$ \citep{auton1988force,hunt1994}, which applies to small spherical bubbles or particles in a shear flow. 


For the purpose of simplification, we will now consider the turbulence to be homogeneous and isotropic. A state of homogeneous isotropic turbulence~(HIT) is fully determined by knowing the kinematic viscosity $\nu$, an outer length scale, and a time scale. For a bubble or light particle rising through HIT~(see  {\bf Figure}~\ref{fig:1_PIT_schematic}(b)), it sees the largest and smallest length scales, which are the energy-injection scale $L$ and the Kolmogorov (or dissipative) scale $\eta$, respectively. Additionally, $\lambda$ represents an intermediate scale known as the Taylor micro-scale, upon which the Taylor Reynolds number Re$_\lambda \equiv u' \lambda/\nu$ is based.
Since the particle size is comparable to the dissipative scale, it is appropriate to non-dimensionalize the equation of particle motion using the Kolmogorov units of length ($\eta$) and time ($\tau_{\eta}$). This leads to 
\begin{equation}
{\bf \ddot{x}_p} = \beta \frac{D\bfu}{D t}   + \frac{1}{\text{St}}(\bfu -  {\bf \dot{x}_p}) + \frac{1}{\text{Fr}} \hat{\bfe}_{y} + \frac{\beta}{3} (\bfu -  {\bf \dot{x}_p}) \times \bm{\omega}, \label{eq_non-dim}
\end{equation}
where the lower bold case variables ${\bf {x}_p}$, ${\bf u}$, and ${\bm \omega}$ denote the new dimensionless vectors for particle position, flow velocity and vorticity, respectively. Here $\beta \equiv {3}/{(1 + 2 \Gamma)}$ is an effective density ratio that takes the fluid added mass into account. The Stokes number $\text{St} \equiv  \frac{d_p^2}{12 \beta \nu \tau_{\eta}}$ and the Froude number $\text{Fr} \equiv  \frac{ a_{\eta}  }{ \left( \beta - 1 \right) g} $
are defined generically to be valid for light ($1 < \beta \leq 3$), heavy ($0 \leq \beta < 1$) and neutrally buoyant~($\beta = 1$) particles. Here $a_\eta \equiv \eta/\tau_\eta^2$ is the acceleration at the Kolmogorov scale.





\section{BUOYANT PARTICLE DYNAMICS}
\label{regimes}

\begin{figure*} [!tbp]
	\centering
	\includegraphics[width=0.85\linewidth]{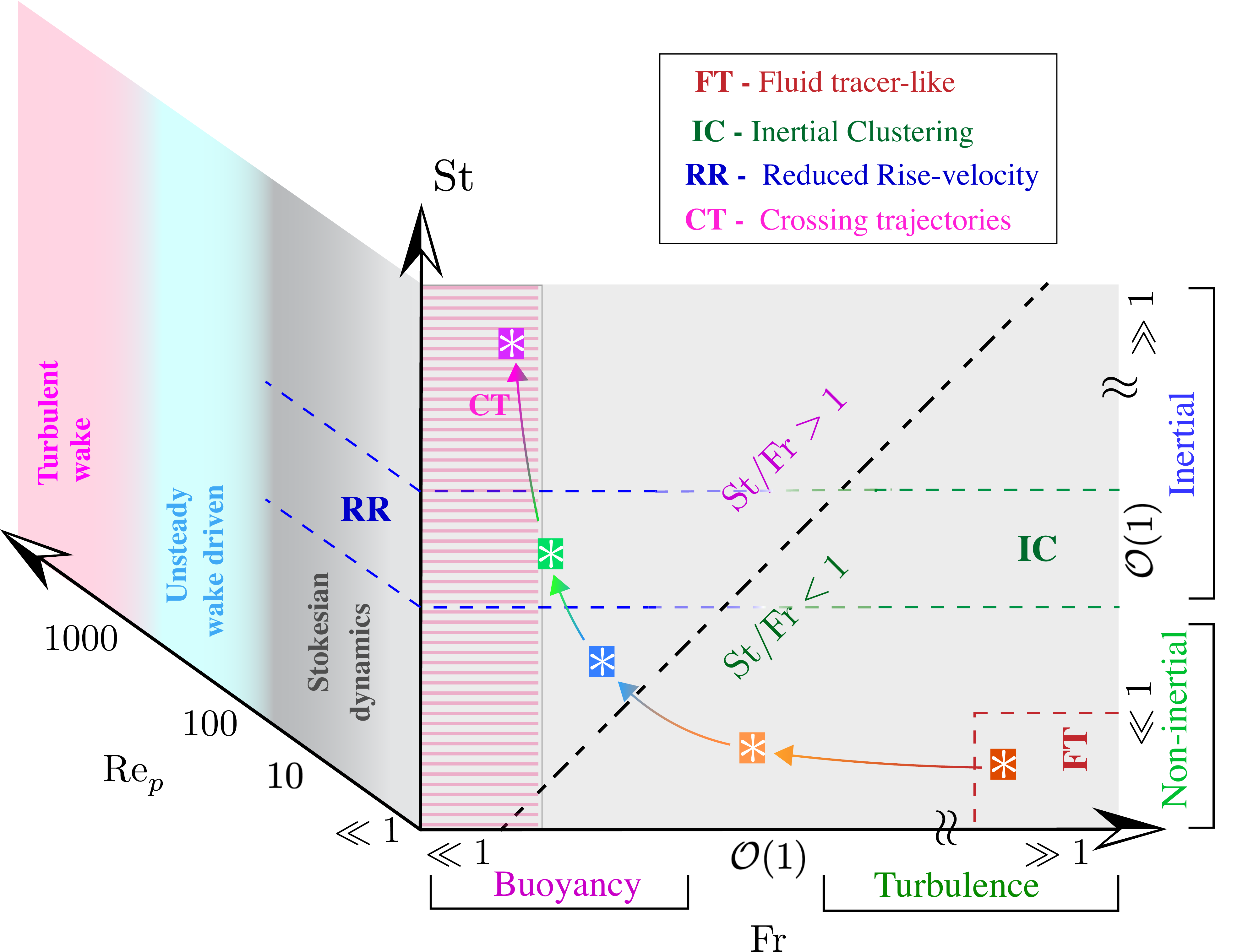}
	\caption{A simplified regime diagram for a buoyant particle (or air bubble) in a turbulent flow. The dynamics of the particle can be summarized as a function of of three dimensionless parameters, namely the Stokes number, St - a measure of particle inertia, the Froude number, Fr - the ratio between turbulence strength and gravity, and the particle Reynolds number, Re$_p$ \citep{calzavarini2009acceleration,mathai2016microbubbles}. The acronyms {\bf FT}, {\bf IC}, {\bf RR}, and {\bf CT} denote the different dynamical regimes, which have been expanded at the top of the figure. The horizontally hatched (magenta) region is the regime where the effect of crossing trajectories can be seen to various extents depending on the value of St. {\color{black} The ${\bf*}$ symbols (connected by arrows) help to picturize a gradual regime transition seen by an air bubble (density ratio $\Gamma = 10^{-3}$) upon increasing its size at constant turbulence level, and assuming that the bubble Reynolds number Re$_p \ll 1$. Note that the effects of changing $\Gamma$ and size ratio ($\Xi$) are not represented in this diagram.}}
	\label{fig:2_regimes}
\end{figure*}

\subsection{Regimes of dynamics}

A wealth of intriguing phenomena have been reported for buoyant particle suspensions in turbulent flows. {\bf Figure}~\ref{fig:2_regimes} outlines a simplified regime diagram for a buoyant spherical particle~(or bubble) when the particle parameters St, Fr, and Re$_p$ are varied, with $\Gamma$, $\Xi$, and background turbulent flow unchanged. When extremely small, a bubble (or light particle) can serve as a passive flow visualization tool~(see {\bf FT} regime in {\bf Figure}~\ref{fig:2_regimes}), while in other scenarios, they have been used to selectively sample intense vortical structures present in flows \citep{douady1991direct,la2001fluid}. In the case of finite particle inertia ($\text{St} \sim \mathcal{O}(1)$ and Fr $\gg$ 1), it is widely known that bubbles cluster in high vorticity regions~(see {\bf IC} regime in {\bf Figure}~\ref{fig:2_regimes}).
\cite{wang1993motion}, \cite{spelt1997motion}, and \cite{mazzitelli2003effect} observed these effects in their simulations of homogeneous isotropic turbulence laden with microbubbles. The phenomenon was later extensively explored by others~\citep{calzavarini2008dimensionality,calzavarini2008quantifying}, thus providing a unified understanding of the clustering behavior of light, heavy and neutrally buoyant particles.  The methods adopted to quantify clustering have been diverse, with \cite{calzavarini2008dimensionality} using the Kaplan-Yorke dimension and Minkowski functionals, while others have used Vorono\"i tessellations  \citep{monchaux2010preferential,tagawa2012three,fiabane2012clustering,obligado2014preferential} or simply the relative particle concentration in high vorticity flow regions \citep{mazzitelli2003relevance}, but all to the same end. An assessment of the actual forces that bubbles are subject to in turbulent environments was performed by \cite{volk2008acceleration,volk2008laser}, who experimentally investigated the acceleration dynamics of small bubbles~($d_p \approx 75$ $\mu$m) in a relatively intense turbulent flow (Re$_\lambda = 850$) generated by a von K\'arm\'an flow apparatus. The high intensity of turbulence ensured that the $75$ $\mu$m air bubble had a St$= 1.85$, with a Froude number Fr $\sim 10$, which meant that the role of buoyancy was negligible. These bubbles showed an acceleration variance four times that of the fluid, which can be presumed to be a combined effect of inertial forces and preferential accumulation in high vorticity regions of the flow \citep{calzavarini2009acceleration}.  For reviews on inertial particle dynamics in the absence of buoyancy, readers are directed to  \cite{toschi2009lagrangian,voth2017anisotropic}. 

{\color{black} It is worthwhile to take a step back to appreciate the various regime transitions seen by an air bubble ($\Gamma = 10^{-3}$) when its size is increased, while keeping all other parameters fixed. The {\bf *} symbols in {\bf Figure}~\ref{fig:2_regimes} show this transition if the bubble Re$_p$ is small.  However, in practicality for a laboratory scale turbulent water flow (say Re$_\lambda \sim \mathcal{O}(100)$ and $L \sim 100$~mm) the actual regime transitions can be more complex. For instance, a microbubble with $d_p$ < 10 $\mu$ m can still be considered a good tracer of the turbulent flow, while upon increasing its diameter $d_p$ to around100 $\mu$m, the effects of buoyancy begin to play a role in the dynamics. When the bubble is a few hundred microns in size, in addition to the buoyancy, the bubble inertia becomes important, and upon increasing the diameter further ($d_p > 500 \mu$m), one can expect noticeable non-Stokesian and finite Re$_p$ contributions. In the following sections, we will discuss these regimes of dynamics in more detail.}

\subsection{Crossing trajectories} 
\label{sec: Crossing_traj}

The dynamics of a particle advected in HIT in a regime where $\text{Fr} \equiv  \frac{ a_{\eta}  }{ \left( \beta - 1 \right) g}  < 1$ will be addressed in this section. The importance of buoyancy (Fr $< 1$) naturally implies that the particle experiences a mean vertical drift through the turbulent flow, and hence the name ``crossing trajectories''~(see {\bf Figure}~\ref{fig:1_PIT_schematic}(b)).

\subsubsection{Non-inertial particles with buoyancy}

\begin{figure*}[!tbp]
	\centering
	\includegraphics[width=0.99\linewidth]{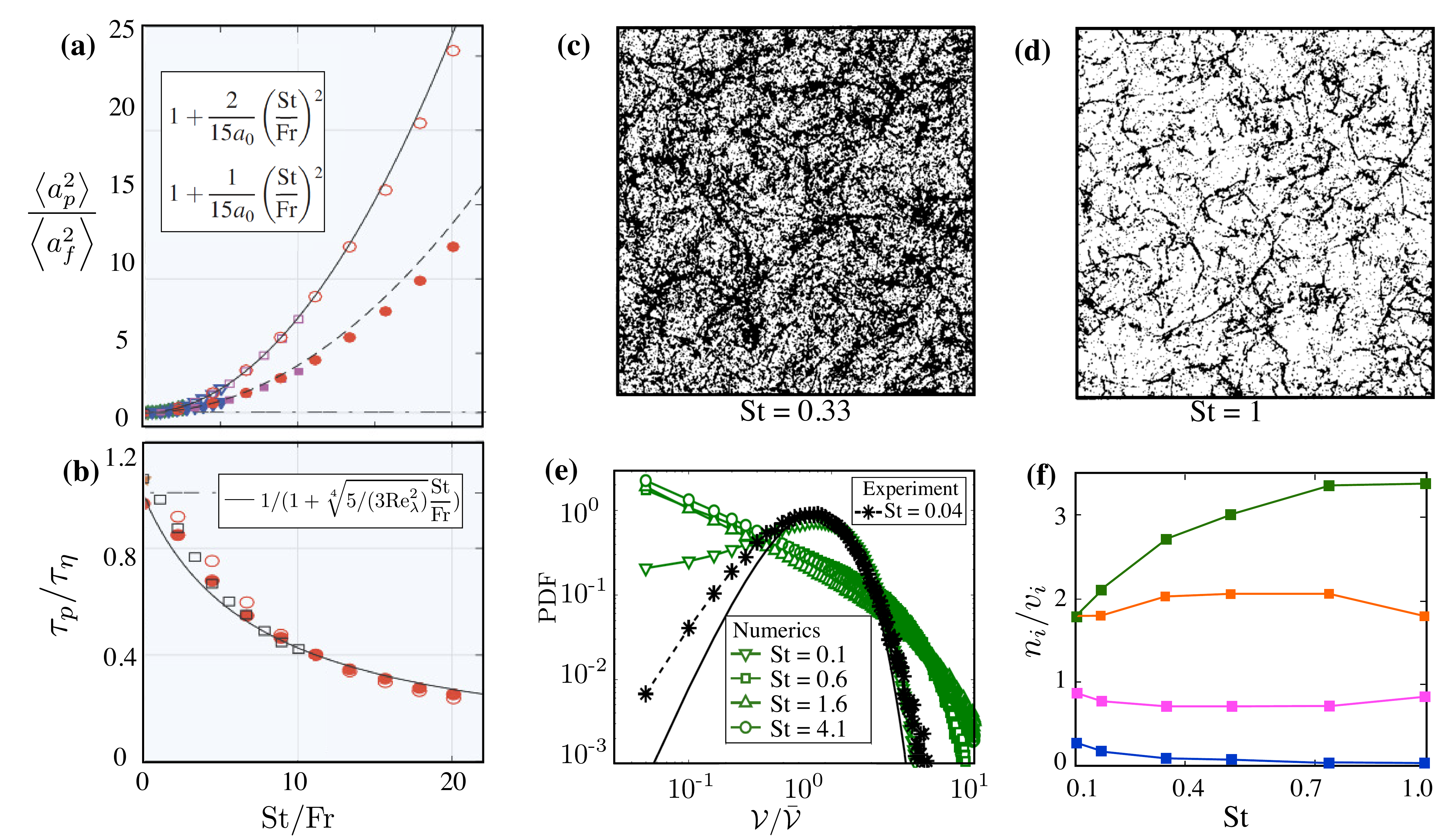}
	\caption{Accelerations and clustering behavior of small ($\Xi  < 1$) buoyant particles and bubbles in isotropic turbulence obtained using Euler-Lagrangian DNS with the inclusion of buoyancy and lift. (a) Normalized acceleration variance of buoyant non-inertial~(St $\ll$ 1) particles in turbulence versus St/Fr at Re$_\lambda = 75$. The acceleration variance here is normalized by the acceleration variance of the fluid. (b) Normalized acceleration decorrelation time~(time to reach an autocorrelation value of 0.5) of the particles for the same cases as in (a). The normalization here is with the Kolmogorov times scale of the turbulence. Solid and hollow symbols in (a) and (b) correspond to vertical~(gravity) and horizontal components, respectively. The solid and dashed curves in (a) and (b) are theoretical predictions. Figures (a,b) adapted from \cite{mathai2016microbubbles}. (c) and (d) Projections of bubble distribution in isotropic turbulence~(Re$_\lambda = 62$) for two Stokes numbers St = 0.33 and St = 1, respectively. Note that the particle concentration is identical for (c) and (d). Bubble clustering is pronounced in the  St  = 1  case (d). We estimated the ratio St/Fr ($\approx v_T/u_\eta$) to be 1.0 and 3.0, respectively, in (c) and (d), which implies R$_v < 1$. Figures adapted from \cite{mazzitelli2003effect}. (e) Clustering of light particles in turbulence, quantified using probability density functions~(PDFs) of normalized Vorono\"i volumes for different values of St. The solid curve is a $\Gamma$-distribution, which is representative for non-clustering (randomly distributed) particles. Green data points are based on DNS using an Euler-Lagrangian (point particle) model, at Re$_\lambda = 180$. Black data points are experimental data for microbubbles~(St = 0.04) in turbulence at Re$_\lambda = 162$. The results suggest that the maximum clustering for bubbles occurs at St around 1-2. This figure was adapted from \cite{tagawa2012three}. (f) Number of bubbles located in a particular zone of fluid $n_i$, normalized by the volume fraction of that zone $v_i$, versus St. Here again Re$_\lambda = 62$, and St/Fr was kept fixed at 1. The symbol colors refer to zones: eddy (green), shear (orange), streaming (magenta) and convergence (blue). This figure is adapted from \cite{mazzitelli2003relevance}. } 
	\label{fig:3_crossing_trajectories}
\end{figure*}

We begin with bubbles and buoyant particles that are in the non-inertial limit, i.e. $\text{St} \ll 1$. Since a very small St naturally implies a tiny particle dimension,
such bubbles are commonly used as tracers in turbulence experiments. Recently, \cite{mathai2016microbubbles} conducted a combined experimental and numerical study on the dynamics of such small bubbles and particles in the non-inertial ($\text{St} \ll1 $) limit. An interesting consequence of buoyancy is that even small St bubbles are subject to intense accelerations. For  Fr $ \ll 1$ and St $\ll 1$, Eq.~(1) is dictated by the balance between just the drag and buoyancy terms. This yields an expression ${\bf \ddot{x}_p} \simeq \frac{D{\bf u}}{Dt} + \frac{\text{St}}{\text{Fr}} {\partial_y{\bf u}}$, where  ${\partial_y{\bf u}}$ are the  gradients of the turbulent flow velocity at the particle's location. The buoyancy parameter can be exactly computed as  R$_v = u_\eta \text{St}/(u' \text{Fr})$.  The acceleration variance ($i^{th}$ component) of the buoyant particle $\left < a_p^2 \right >_i $ can be expressed as an enhancement over the fluid acceleration variance $\left < a_f^2  \right > $:

\begin{equation}
\frac{\left < a_p^2 \right >_i}{\left <a_{{f}}^2 \right >}  \simeq 1 +  \kappa_i \left( \frac{\text{St}}{\text{Fr}}\right)^2,
\label{eq_avar_non_inertial}
\end{equation}
where $\kappa_x = 2/(15 a_0)$ for the horizontal component, and  $\kappa_y  = 1/(15a_0)$ for the vertical component, with $a_0$ the so-called Heisenberg-Yaglom constant \citep{voth2002measurement}. These relations follow exactly from the assumption that the turbulence is statistically isotropic \citep{pop00}. In other words, the increase in acceleration variance is a direct consequence of the vertical drift of the bubble through the turbulent eddies. While the acceleration variance increases, its decorrelation time compared to the fluid is suppressed~(see {\bf Figure}~\ref{fig:3_crossing_trajectories}(a)-(b)), as the drifting particle spends comparatively less time within the turbulent eddies. Similarly, the kurtosis of acceleration is diminished, a consequence of the fact that the spatial velocity gradients $\partial_y {\bf u}$ in turbulence are less intermittent than the fluid acceleration~(for further details, see \cite{mathai2016microbubbles}). Finally, we note that in this non-inertial limit, the behavior of buoyant particles is expected to be nearly identical to that of heavy particles, for fixed St/Fr~\citep{csanady1963turbulent,maxey1987gravitational,parishani2015effects}. Of course, the effect is more pronounced for bubbles as their density contrast is generally larger as compared to that of the heavy particles in liquid flows. Thus, a tiny bubble or droplet is not necessarily a good tracer of turbulent acceleration. In  reality, the situation of finite $g/a_\eta$ is common for bubbles that drift through the turbulent oceans
($g/a_\eta \approx  100-1000$), and for droplets settling through clouds ($g/a_\eta \approx 10-100$) \citep{bodenschatz2010can}. On
the practical side, these also point to the key condition $\text{St/Fr} \ll 1$ that must be met (in addition to St $\ll 1$) for the usage of small bubbles (or droplets) for flow visualization and particle tracking in turbulent flows.


\begin{figure*} [!htbp]
	\centering
	\includegraphics[width=0.99\linewidth]{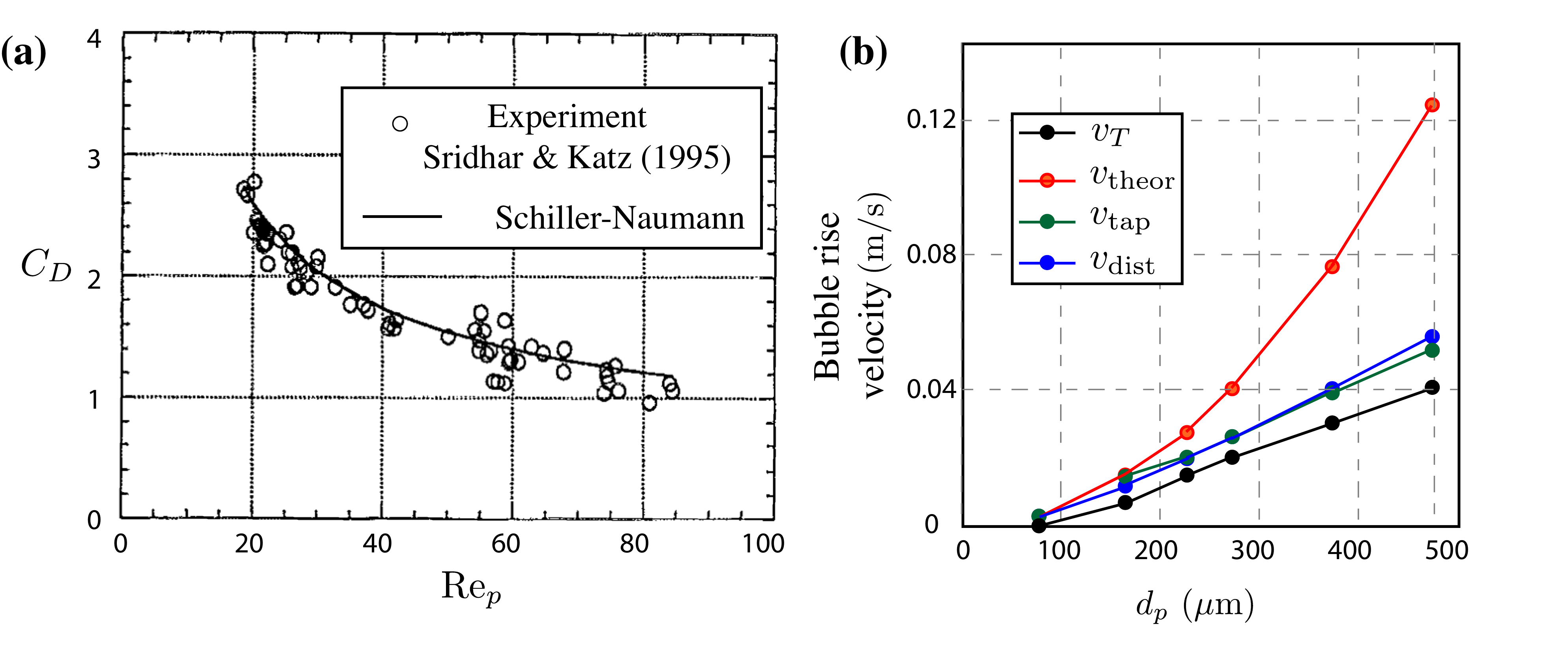}
	\caption{Drag coefficient and rise velocity modifications for sub-millimetric air bubbles rising in turbulent flows. (a) Comparison of the drag coefficient for $500 - 800~\mu$m diameter air bubbles (in a vortical flow), along with the predictions of the Schiller-Naumann drag model. Data reproduced from \cite{sridhar1995drag}. (b) Direct measurements of rise velocity $v_T$ for small bubbles versus bubble diameter $d_p$, in a grid generated turbulent flow at Re$_\lambda \approx 404$~(as roughly estimated by us). For comparison the following are given. $v_\text{theor}$: the terminal velocity estimated theoretically using Stokes drag $C_D = 24/$Re$_p$, $v_\text{dist}$: the terminal velocity measured in triply distilled water, and $v_\text{tap}$: the same quantity measured for tap water. Data adapted from \cite{aliseda2011preferential}.}
	\label{fig:4_reduced_rise_drag}
\end{figure*}

\subsubsection{Inertia and buoyancy}
\label{sec: inertia_buoyancy}

Next, we consider the cases when Fr $\leq1$ and St $\sim 1$, i.e. a regime where both buoyancy and inertia are  important, and in competition. Employing an Euler-Lagrangian point particle (PP) approach of a form similar to eq~(\ref{eq_non-dim}), \cite{mazzitelli2003effect} studied the effect of increasing St at fixed Fr for bubbles rising in isotropic turbulence~(Re$_\lambda = 62$). Snapshots of the bubble distribution obtained from their simulations are shown in {\bf Figure}~\ref{fig:3_crossing_trajectories}(c)-(d). At St = 0.33, the particles~(bubbles) are nearly uniformly distributed, but at St = 1  clustering is visibly amplified. 
Three-dimensional Vorono\"i analysis was used to quantify the clustering of bubbles in homogeneous isotropic turbulence using data sets from numerics in the point particle limit (without gravity and lift) and an experimental data \citep{tagawa2012three}. In the PDFs of normalized Vorono\"i volumes ($\mathcal{V/\overline{V}}$) shown in {\bf figure} \ref{fig:3_crossing_trajectories}(e), the solid black curve - a $\Gamma$-distribution - is representative of non-clustering (randomly distributed) particles. 
For bubbles, they observed that the probability of finding both small and large Vorono\"i volumes $\mathcal{V/\overline{V}}$ is generally higher. The two regions of small and large volumes can be used to identify clusters and voids. A high probability for low values of $\mathcal{V/\overline{V}}$  is signatory of intense clustering. As shown in the figure, when St increases, the probability of finding clusters (and voids) increases, and reaches a maximum value at  St $\sim$ 1.6, suggesting that the strongest clustering for bubbles occurs in the St range  1-2. 
A direct quantification of the clustering in the presence of buoyancy is shown in {\bf Figure}~\ref{fig:3_crossing_trajectories}(f), where the number fraction of bubbles located in a vorticity dominated region is given by the green symbols. 
Since the buoyancy and lift terms were included by \cite{mazzitelli2003relevance}, the relative degree of clustering is typically less than what is seen in minimalistic simulations~\citep{calzavarini2008quantifying} where these terms were neglected.

\subsubsection{Non-Stokesian bubbles and particles}
For the more commonly encountered situation of  air bubbles in water flows, a Stokes number of order one typically almost never satisfies the condition of Re$_p \ll 1$~\citep{magnaudet2000motion,mathai2018dispersion}. Hence a modified consideration of the drag,
added mass and lift forces is essential to predict the trajectory of finite sized and finite Reynolds number bubbles. As a model problem, \cite{sridhar1995drag}  experimentally studied the entrainment of such air bubbles~($d_p \approx 500 - 800$ $\mu$m) by a vortex ring. They reported that the drag coefficient was comparable to the Schiller-Naumann parameterization (see {\bf Figure}~\ref{fig:4_reduced_rise_drag}(a)), while the lift forces did not agree with the existing theoretical or numerical models~ \citep{maxey1983equation,maxey1987gravitational,tio1993dynamics}. More recently, \cite{aliseda2011preferential} experimentally investigated the behavior of small spherical bubbles~(100 -- 1000 $\mu$m) immersed in a homogeneous isotropic turbulent water flow (see {\bf Figure}~\ref{fig:4_reduced_rise_drag}(b)). Within the turbulent flow the concentration field of the bubbles was altered, with preferential accumulation at the smaller scales and reduced rise velocities as compared to the value in stationary liquid. These can be interpreted as occurring due to two phenomena. Firstly, the pressure fluctuations drive the inertial bubbles to the cores of the vortices. Secondly, the lift forces cause the bubbles to be preferentially transported toward downflow regions, where, in combination with an increased relative velocity (increased viscous drag), they are further slowed down~(see cartoon in {\bf Figure}~\ref{fig:1_PIT_schematic}(a)) \citep{mazzitelli2003relevance}.

\begin{figure*} [!tbp]
	\centering
	\includegraphics[width=0.99\linewidth]{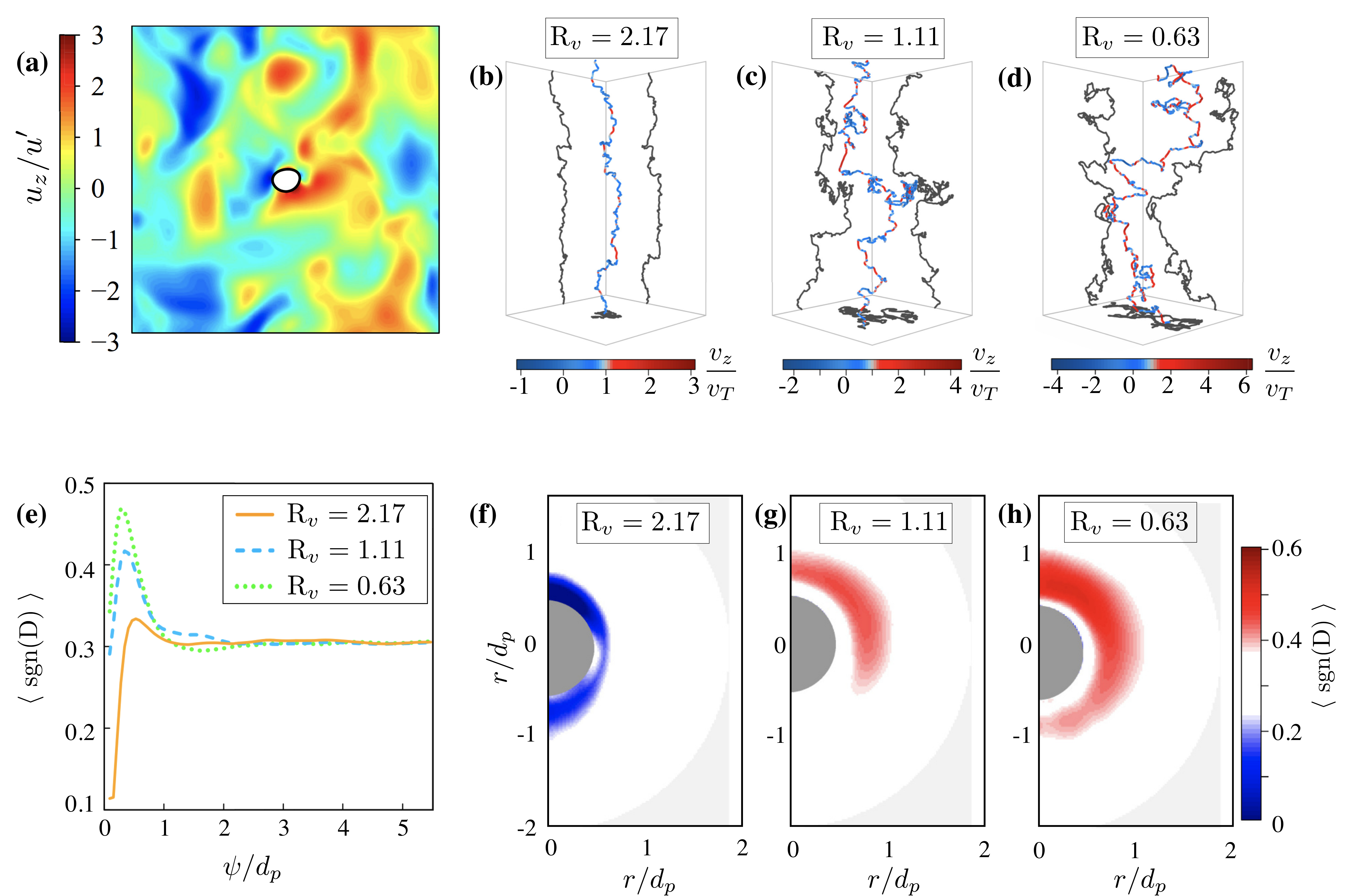}
	\caption{Direct numerical simulations of a slightly deformed bubble~($\Xi \approx 10$, Re$_p \in [17, 62]$) rising in a turbulent flow at Re$_{\lambda} = 30$. (a) Snapshot of the (normalized) vertical velocity field of the turbulent flow on a plane passing through the bubble center. (b)-(d) Sample bubble trajectories for decreasing buoyancy parameter R$_v \equiv v_T/u'$. R$_v = 2.17$ in (b), 1.11 in (c) and 0.63 in (d). The bubble trajectories are colored by their instantaneous vertical velocity normalized by their terminal velocity. (e) Radially averaged profile of discriminant D around the bubble for the three different R$_v$ cases. The abscissa $\psi/d_p$ is the (normalized) radial distance from the centroid of the bubble. The sign of the discriminant D (ordinate) can be used to highlight the vorticity-dominated regions. (f)-(h) Average discriminant field around the bubble for decreasing R$_v$ (left to right). Color scale centered around the mean value $\overline{\text{D}} \approx 3$ to show that values above $\overline{\text{D}}$ are the vorticity dominated regions. The discriminant field clearly demonstrates that clustering in high vorticity regions is increased as the bubble rise velocity decreases.  Data and figures adapted from \cite{loisy2017interaction}.}
	\label{fig:5_Loisy_figures}
\end{figure*}

The conditions of small size ($d_p/\eta \leq 1$) and nearly spherical shape is typically satisfied only for sub-millimeter air bubbles \citep{aliseda2011preferential,sridhar1995drag}. However, the vast majority of bubble laden turbulence operates under conditions where the bubbles are of finite size, free to move, and, most importantly, deformable  \citep{deane2002scale,duineveld1995rise}. These call for a more detailed consideration of the coupled interaction between the bubble topology and its wake-induced dynamics arising from the finite Re$_p$ and finite Weber number effects \citep{ryskin1984numerical}. 
A fully resolved treatment of the interaction between isotropic turbulence and large solid spherical
particles has been performed in a variety of configurations \citep{naso2010interaction,homann2010finite,chouippe2015forcing}.  In comparison, DNS of
turbulent bubbly flows are challenging, owing to the interface deformations and internal circulations, along with the need to resolve a wide range of length and times scales inherent to the turbulent flow. \cite{loisy2017interaction} studied an isolated deformable bubble freely rising in an otherwise isotropic turbulent flow using DNS~(see  {\bf Figure}~\ref{fig:5_Loisy_figures}(a)). The buoyancy parameter (R$_v \in [0.63, 2.17]$) and bubble Reynolds number~(Re$_p \in$ [17, 62]) were both moderate, and hence in quiescent liquid the bubble rises along straight vertical paths. However, with decreasing R$_v$ the trajectories become erratic and increasingly deviate from vertical paths~(see  {\bf Figure}~\ref{fig:5_Loisy_figures}(b)-(d)). This was accompanied by a reduction
in rise velocities. With regard to the statistics of bubble velocity and acceleration, the probability density functions~(PDFs) was nearly Gaussian for the velocity and showed stretched tails for accelerations. Lastly, the bubble showed a preference for increased residence in vorticity-dominated regions, here, computed using the discriminant $D = 27R^2 + 4Q^3$, where $Q$ and $R$ are the second and third invariants of the velocity gradient tensor, respectively  \citep{naso2005scale}. This was revealed through conditional sampling of the average discriminant profile and discriminant field around the bubble (see  {\bf Figure}~\ref{fig:5_Loisy_figures}(e) and {\bf Figure}~\ref{fig:5_Loisy_figures}(f)-(h), respectively). 

The analyses of \cite{loisy2017interaction} (given above)  show that the dynamics of a moderate Re$_p$ bubble in turbulence (rise velocity, PDF shapes and clustering) are, at least, qualitatively captured by the point-particle model \citep{calzavarini2009acceleration}, despite their finite Re$_p$, finite size, and deformed shape. However, this is not reflective of the full picture, since certain aspects of the bubble statistics are markedly different from the PP predictions. The PDF of the longitudinal acceleration, i.e. the component of bubble acceleration directed along its instantaneous velocity, was found to be negatively skewed, a feature not captured by the PP model even with the inclusion of a back reaction force. Whether the origin of this lies truly in the  time-irreversibility \citep{xu2014flight} of turbulence (as postulated by \cite{loisy2017interaction}), or in the asymmetry of flow induced forcing on the bubble, remains to be unravelled.

%


\label{sec-mean_flow}

\subsection{Wake-driven dynamics and path-instabilities} 
\label{sec-wakedriven}

With increasing buoyancy over inertia, the bubble or buoyant particle's Reynolds number Re$_p$ can reach a few hundreds. Two important changes come into effect in this regime. Firstly, the mean drag coefficient \textcolor{black}{loses its Reynolds number dependence, and $C_D$ becomes weakly dependent on Re$_p$.} In addition, such bubbles and particles also experience fluctuating components of forces that originate from the instability of their wakes. Although the mean forces on the particle can still be approximated, the instantaneous drag and lift can no longer be described using simplified coefficients. Furthermore, owing to the lightness of the particle, this regime paves way for a strong coupling between the unsteady wake dynamics and the particle motion, often resulting in vigorous path instabilities~\citep{ern2012wake,mougin2006wake,mathai2016translational,mathai2018flutter,brucker1999structure, mathai2017mass}. 
As reported in \cite{risso2017agitation}, there is, today, compelling evidence that the wakes and dynamics of isolated buoyant particles are remarkably robust to turbulent perturbations \citep{ford1998forces}. Therefore, the forces acting on an isolated buoyant particle in a flow can still provide a basis for understanding dispersed two phase flows.

\begin{figure*} [!tbp]
	\centering
	\includegraphics[width=0.99\linewidth]{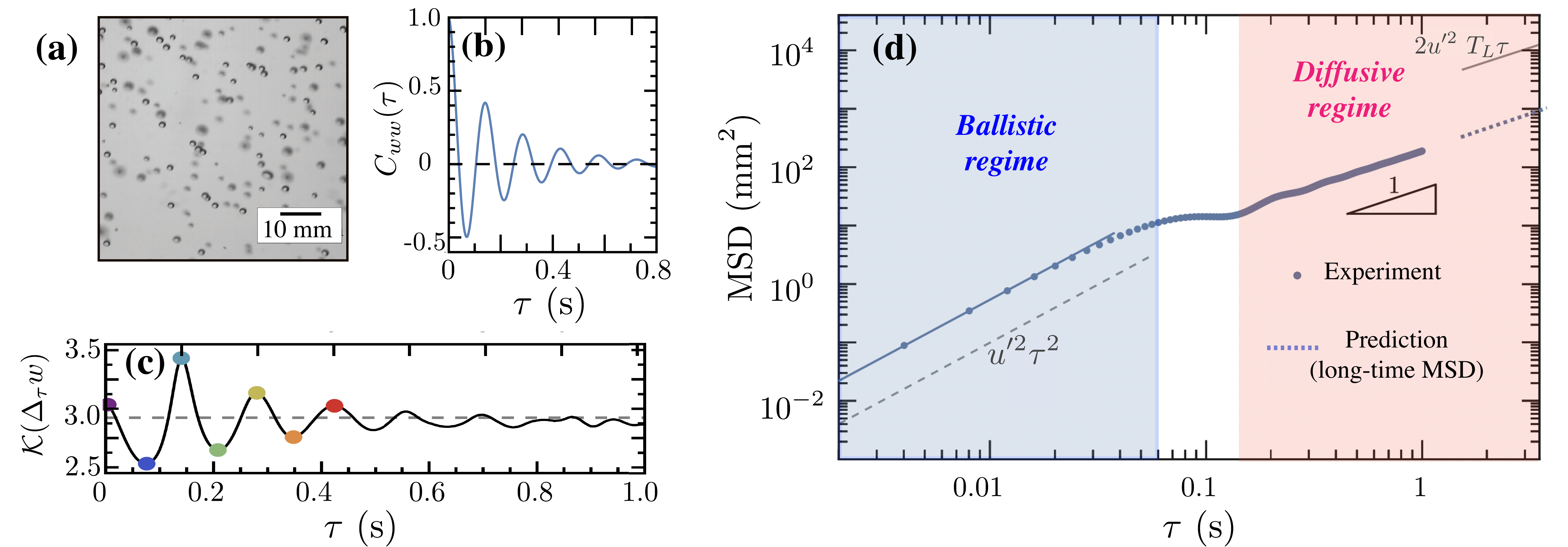}
	\caption{The horizontal component of velocity and dispersion  for 1.8 mm diameter air bubbles~($R_v \approx$ 10) rising in active-grid-generated turbulence at Re$_\lambda$ = 110. Here the bubble size ratio $\Xi \approx 5$. (a) A sample image from one of the cameras.  (b) Lagrangian temporal autocorrelation function of the horizontal component of the bubble velocity. (c) Kurtosis of the horizontal velocity-increment PDFs as a function of time lag $\tau$. (d) Horizontal component of the mean squared displacement~(MSD) for the bubble as a function of time lag $\tau$. The well-known {\it ballistic-diffusive} behavior of fluid tracers is shown for comparison: dashed gray line ($\propto \tau^2$), and solid gray line ($\propto 2T_L \tau$). The dotted blue line on the right shows the prediction for the reduced long-time dispersion for the bubble, obtained using a crossing-trajectory hypothesis. Data and figures adapted from \cite{mathai2018dispersion}.}
	\label{fig:6_Bubble_dispersion}
\end{figure*}


\subsubsection{Wake-driven bubbles}

Building on the original formulations proposed by Kirchhoff  \citep{lamb1993hydrodynamics,galper1995dynamic} for a rigid particle in a arbitrary irrotational flow,  \cite{mougin2001path} extended the case to a situation where the NS equations governing the liquid flow \citep{mougin2002generalized,auguste2017path} are considered in conjunction with Newton's laws for the translational and rotational dynamics of a ``fixed-in-shape'' non-spherical bubble or buoyant particle. The Kelvin-Kirchhoff equations, which disregard small-scale deformability effects, demonstrated that wake instability and anisotropic added-mass effects of oblate spheroids~\citep{cano2016paths,mougin2006wake} are indeed sufficient to explain the experimentally observed path instabilities \citep{wu2002experimental} of millimetric bubbles. 
The method, now at our fingertips, is yet to be used widely for turbulent bubble laden flows, and holds the potential to yield useful insights at a reduced computational cost, since the condition of a continuously deforming bubble interface is relaxed. 

Experimentally, the dynamics of isolated millimetric air bubbles in turbulent flows was studied by \cite{mathai2018dispersion}. In this work, the authors generated a suspension of 1.8 mm diameter air bubbles in an active grid generated homogeneous isotropic turbulent water flow with Re$_\lambda \in [110, 300]$. The bubbles were nearly spherical~(see  {\bf Figure}~\ref{fig:6_Bubble_dispersion}(a)), and their volume fraction in the experiments was low~($\alpha \sim 5 \times10^{-4}$). At a low level of turbulence~(Re$_\lambda = 110$), the Lagrangian temporal autocorrelation of the bubble velocity and its kurtosis both showed periodicity~(see  {\bf Figure}~\ref{fig:6_Bubble_dispersion}(b)-(c)), clearly indicative of wake shedding at a frequency $f_{\text{viv}} \sim 0.1 v_p/d_p$. The effects of these on the spreading of the bubbles was analyzed using the mean-squared-displacement~(MSD), which was then compared to the well-known regimes of Taylor-dispersion for the fluid in turbulence~(see {\bf Figure}~\ref{fig:6_Bubble_dispersion}(d)). At short
times, the bubble MSD grows ballistically ($\propto \tau^2$), whereas at a larger time scale set by the wake-shedding frequency $f_\text{viv}$, it approaches the diffusive regime where the
MSD $\propto \tau$. Note that for the bubbles, the ballistic regime lies well above the $u'^2 \tau^2$ prediction of fluid tracers, while the diffusive regime MSD for the bubbles lies well below the $2 u'^2 T_L \tau$ of fluid tracers. Here $T_L$ is the Lagrangian integral time scale which sets the ballistic-diffusive transition time for fluid tracers in turbulence. 
To conclude, with high Reynolds number millimetric air bubbles in turbulence, we can appreciate an elegant merger of two classical phenomena, namely the wake-induced velocity fluctuations of the bubbles (at short times), and the reduced dispersion (at longer times) originating from the crossing trajectories effect \citep{calzavarini2018propelled, mazzitelli2004lagrangian,mathai2018dispersion,mathai2016microbubbles}.

\begin{figure*} [!tbp]
	\centering
	\includegraphics[width=0.9\linewidth]{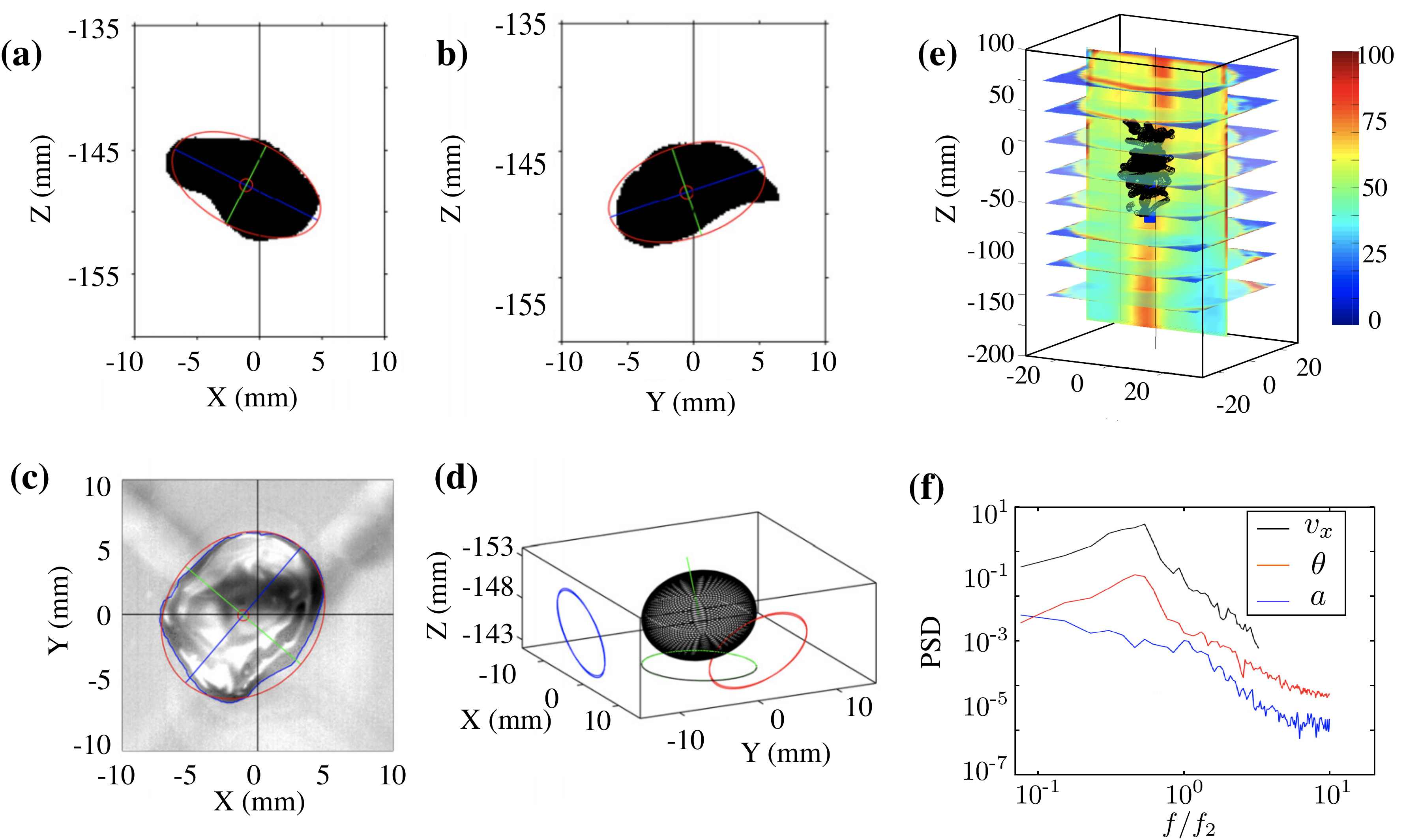}
	\caption{Kinematics of a deformable air bubble~($d_p \approx 9.3$ mm, Re$_p$ = 2800) rising in a (rotating) grid-generated turbulent flow, based on \cite{ravelet2011dynamics}. The buoyancy parameter R$_v$ lies in the range [5, 10]. (a)-(d)~Illustration of image processing steps on the bubble images, with equivalent ellipse~(red), center of mass~(red circle), major axis~(blue) and minor~(green) axis  of the best-fit ellipsoid. (e) Standard deviation of the liquid velocity measured in one vertical plane and eight horizontal planes, superimposed with a trajectory of the bubble (black). Note that the bubble here is trapped in a region of highest vorticity. (f) Normalized power spectra of horizontal velocity $v_x$ (black), orientation $\theta$ (red), and major axis $a$~(blue). The spectra are normalized by the frequency $f_2  = 15$~Hz, as done by \cite{ravelet2011dynamics}.}
	\label{fig:7_RaveletRisso}
\end{figure*}

The motion of an even larger bubble~($d_p \approx 9$ mm) in turbulence was considered by \cite{ravelet2011dynamics}. At these sizes in a turbulent water flow, the bubble shows considerable deformability (We $\simeq 11.6$), and its Reynolds number $Re_p \approx 2800$. The researchers tracked the bubble motion and orientation in intricate detail using three-dimensional shape recognition, yielding statistics of bubble translation, rotation and deformation in the turbulence~(see  {\bf Figure}~\ref{fig:7_RaveletRisso}(a)-(d)). The bubble dynamics was found to be governed by three fairly independent mechanisms. The average bubble shape is imposed by the mean motion of the bubble relative to the liquid. Further to this, wake instability of the bubble generates periodic oscillations in its velocity and orientation. Lastly, the turbulence adds to the random deformations, which under rare circumstances can even lead to bubble breakup. The bubbles were observed to be trapped inside a vortex at the core of the flow~(see  {\bf Figure}~\ref{fig:7_RaveletRisso}(e)). The temporal spectra of horizontal velocity, orientation and semi-axis lengths~({\bf Figure}~\ref{fig:7_RaveletRisso}(f)) reveal peaks at around 8.5 Hz. This is consistent with a wake instability and yields a Strouhal number $\text{St} \approx  0.27$. 

\begin{figure*} [!htbp]
	\centering
	\includegraphics[width=0.99\linewidth]{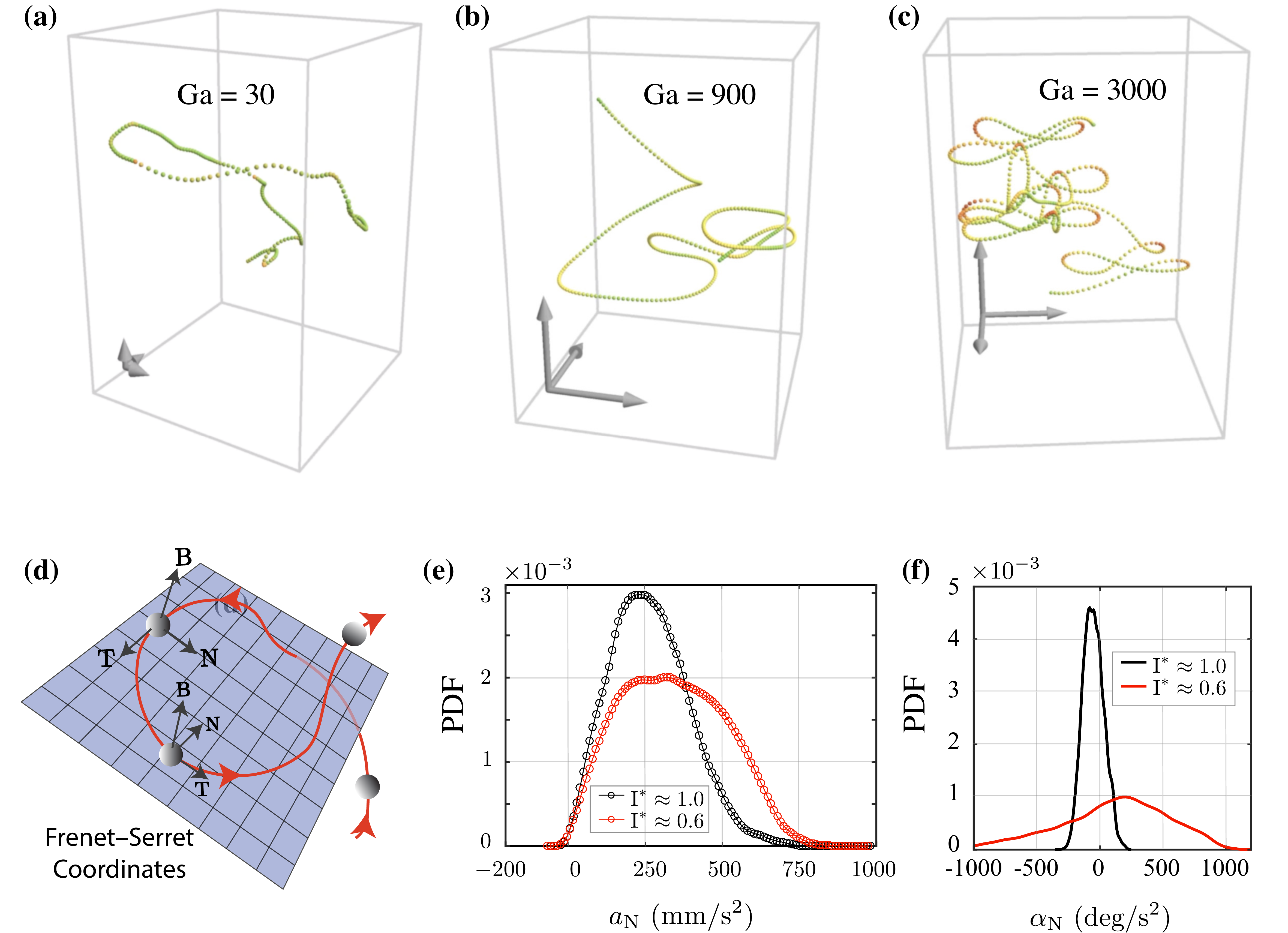}
	\caption{Kinematics and dynamics of rigid buoyant spheres~($\Xi \approx 10$) in a turbulent flow with Re$_\lambda \approx 300$. (a)-(c) Three dimensional trajectories of buoyant spheres for increasing Galileo number, Ga. In obtaining these trajectories, the mean rising motion of the spheres were counteracted by matching the  sphere rise velocity with the mean downward  flow speed. Note that the Ga increase~(in (a)-(c)) coincides with an increase in the buoyancy parameter R$_v \equiv \frac{v_T}{u'}$, which was measured to be  around 1.7 in (a), 13.6 in (b) and 45.6 in (c). (d) The Frenet-Serret coordinates: {\bf TNB},  used to decompose the instantaneous motion of the sphere in turbulence. (e) PDFs of the centripetal acceleration $a_\text{N}$ for two spheres with identical properties ($\Gamma= 0.89$; $\Xi \approx$  100, Ga = 6000), except their mass moments of inertia I$^* = $ 1.0 (black circles) and 0.6 (red circles). Here $a_\text{N}$ is directed along the {\bf N} vector in the Frenet-Serret formulae. (e) Reducing I$^*$ triggers a dramatic increase in the translational and rotational accelerations~(see (e) and (f)). Drawings and figures are adapted from \cite{mathai2015wake,mathai2018flutter}.}
	\label{fig:8_mathai_wake}
\end{figure*}

\subsubsection{Wake-driven rigid buoyant particles}
In the following discussion we provide an overview of studies on large buoyant particles of fixed shape advected in turbulent flows. Initial headway was made using finite sized neutrally buoyant spherical particles  \citep{zimmermann2011tracking,zimmermann2011rotational,bellani2012slip}. From a modeling perspective, the point-particle approach with the so-called Fa\'xen corrections  \citep{faxen1922widerstand,maxey1983equation} for particle size was adopted widely to study neutrally buoyant spherical particles  \citep{calzavarini2009acceleration}. Still, the question of finite slip velocities, which occur in most practical situations, raises important concerns about the validity of this treatment  \citep{prosperetti2015life}. By performing experiments using marginally buoyant~($\beta \approx 1.05$) rigid finite sized spheres in turbulence, \cite{mathai2015wake} showed that even a slight density mismatch is sufficient to cause the dynamics of buoyant particles to deviate significantly from the Fa\'xen model predictions. The deviations increase dramatically upon reducing the density ratio, until for very buoyant particles with $\beta \approx 2.90$, i.e. comparable to the $\beta$ of a bubble, vigorous path oscillations outweigh the turbulence induced motions. {\bf Figure}~\ref{fig:8_mathai_wake}(a)-(c) demonstrate this effect of increasing buoyancy (Ga from 30 to 3000), whereupon the turbulence induced chaotic dynamics (in (a)) are replaced by wake induced oscillations~(in (b) and (c)) reminiscent of Lissajous orbits \citep{govardhan2005vortex}. Three mechanisms contribute to the increasing path-oscillations witnessed here. Firstly, a Ga increase makes the wake-induced forces stronger. Next, an accompanying increase in the buoyancy parameter R$_v \equiv v_T/u'$ occurs. When R$_v$ increases, the rising spherical particle crosses the turbulent eddies at increasingly higher speeds, thus having little time to respond to the turbulent fluctuations. 
A third, not so obvious, influence was revealed experimentally in a recent study by \cite{mathai2018flutter}, which hints that the observed path-instabilities are augmented by the particle's rotational motions as well. To explain this, we revisit the Kelvin-Kirchhoff equations expressing linear and angular momentum conservation for a buoyant spherical particle:
\begin{eqnarray}
({\Gamma} + \frac{1}{2} + \text{B}_1 \ \delta)\ \frac{\text{d}{\bf V}_p}{\text{dt}} +  \Gamma \ {\bm{\Omega}_p \times {\bf V}_p}   =  \frac{\textbf{F}_{Q}}{\rho_l \mathcal{V}_p} + (\Gamma - 1) g {\bf \hat{e}}_j ;
\label{forceeqn}
\end{eqnarray}
\begin{equation}
(\frac{1}{10} \text{I}^* +\text{B}_2 \ \delta) \frac{\text{d}{ \bm{\Omega}}_p}{\text{dt}} =  \frac{{\bf T}_Q}{\rho_p \mathcal{V}_p d_p^2};
\label{momenteqn}
\end{equation}
where ${\bf V}_p$ is the sphere velocity vector, ${\bm{\Omega}_p}$ is the sphere angular velocity vector,  $g$ is the acceleration due to gravity, I$^* \equiv \text{I}_p/\text{I}_l$ is the moment of inertia ratio, with I$_p$ the sphere moment of inertia, and I$_l$ the moment of inertia of the liquid volume displaced by the spherical particle. ${\bf F}_Q$ and ${\bf T}_Q$  represent the ``quasi-static'' fluid force and torque vectors, respectively, which result from the existence of vorticity in the flow. These terms can be straightforwardly obtained by integrating the local stress and moment over the sphere surface. Note that $\delta = \sqrt{\frac{\nu \tau_v}{\pi d_p^2}}$ is the dimensionless Stokes boundary layer which develops in a time $\tau_v$. The prefactors $\text{B}_1 = 18$ and $\text{B}_2 = 2$ are analytically obtained from unsteady viscous contributions~ \citep{zhang1998oscillatory,auguste2017path}.
Equations (\ref{forceeqn}) \& (\ref{momenteqn}) help appreciate the strong coupling that could exist between translation and rotation for a buoyant spherical particle~(I$^* \sim \Gamma < 1$). Assuming the time available for the Stokes layer to develop scales with the vortex shedding time scale, \cite{mathai2018flutter} estimated that up to a moderate Ga, the role of I$^*$ ought to be insignificant. Whereas, upon increasing Ga further, I$^*$ becomes increasingly dominant in Eq.~(\ref{momenteqn}). 

\begin{figure*} [!tbp]
	\centering
	\includegraphics[width=0.7\linewidth]{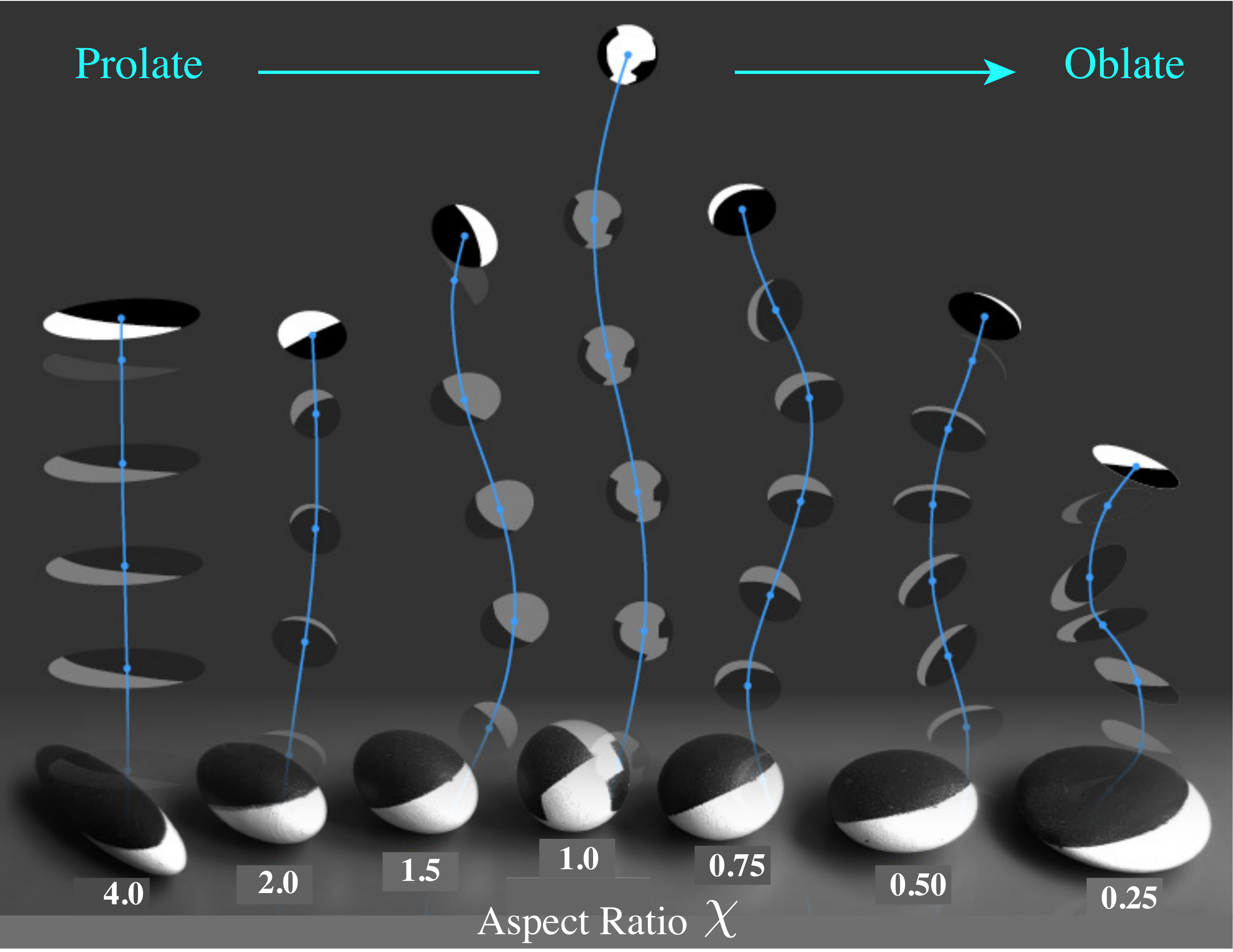}
	\caption{Trajectories of buoyant ellipsoids of different aspect ratios $\chi$  rising in quiescent liquid. The trajectories seen are remarkably robust to background turbulent fluctuations. Here, $\Gamma = 0.8$, and Ga = 300 calculated based on the diameter $d_{pe} = (6 \mathcal{V}_p/\pi)^{1/3}$. $\chi$ is used to distinguish oblate~($\chi <1$) and prolate ($\chi >1$)  ellipsoids from a spherical particle ($\chi =1$). The oblate ellipsoids provides an approximation of the mean deformed shapes  of millimetric air bubbles in water \citep{zenit2008path,ravelet2011dynamics}. Figures based on \cite{will2018large} (unpublished).}
	\label{fig:9_Anisotropic}
\end{figure*}

\subsubsection{Rotation induced accelerations}
To analyse the role of rotation further, it is advantageous to adopt a Lagrangian frame of reference that is oriented with respect to the sphere's instantaneous motion~(see  {\bf Figure}~\ref{fig:8_mathai_wake}(d)). The mutually orthogonal Frenet-Serret coordinates are  the tangent ${\bf T} = {\dot{\bf x_p}}/|{\dot{\bf x_p}}|$, normal ${\bf {N}} = {{\bf B} \times {\bf T}}$, and binormal ${\bf {B}} = ({\bf \dot{x}_p}\times {\bf \ddot{x}_p}) / \mathcal{k}  {\bf \dot{x}_p} \times  {\bf \dot{x}_p} \mathcal{k} $ vectors, which are directed along the particle velocity, curvature, and a direction perpendicular to the trajectory plane, respectively.  For neutrally buoyant spheres in turbulence, \cite{zimmermann2011rotational} had originally shown the existence of an alignment between translation and rotation. {\bf Figure}~\ref{fig:8_mathai_wake}(e) and inset show PDFs of linear~($a_\text{N}$) and angular accelerations~($\alpha_\text{N}$), respectively, of two buoyant spheres~($\Gamma \approx 0.89$) that differ solely in their rotational inertia ratios $\text{I}^*=$ 1.0 and 0.6. For the lower I$^*$ case, a strong coupling between translation and rotation ensues, which reflects strongly in the particle's linear and angular acceleration PDFs  \citep{mathai2018flutter}. The same qualitative effects were reproduced for buoyant cylindrical particles having different I$^*$ inertia  \citep{mathai2017mass}.  {\color{black} New experiments are being extended to the realm of buoyant ellipsoidal particles (oblate to prolate) in turbulence  \citep{will2018large}. Interesting new regimes are being revealed~(see {\bf Figure}~\ref{fig:9_Anisotropic}) due to the coupling between particle buoyancy, particle shape, and turbulence.}

\subsection{Lift and shear-induced lateral migrations}
\label{lift_lateral}

\begin{figure*} [!tbp]
	\centering
	\includegraphics[width=0.78\linewidth]{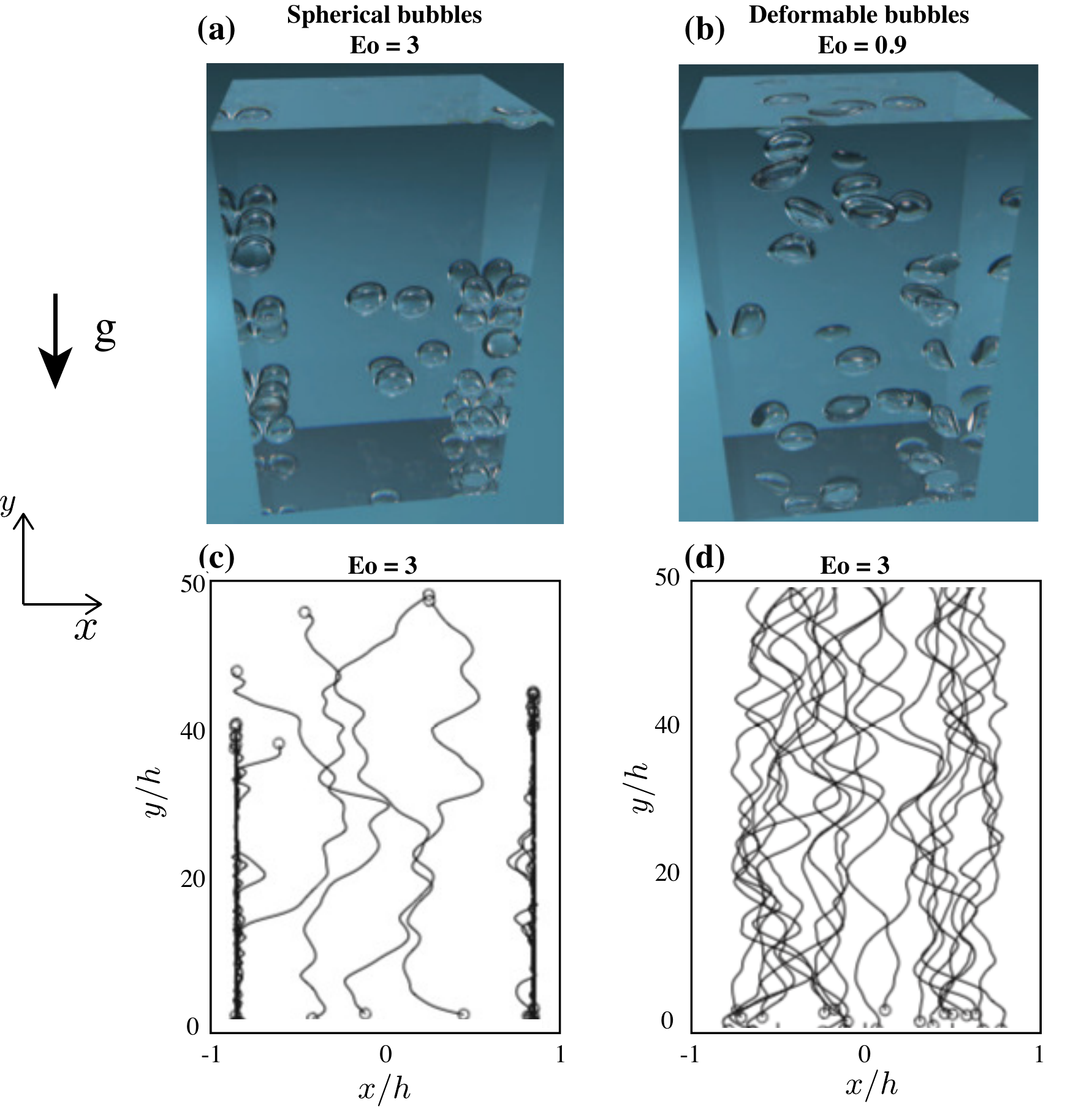}
	\caption{Direct numerical simulations of rising bubbles in vertical channel flows. (a) and (b) Distribution of spherical~(Eo = 3) and deformable bubbles~(Eo = 0.9), respectively, in an upward flow turbulent channel at a friction Reynolds number Re$_\tau = 127$. The ratio of bubble diameter to channel halfwidth $d_p/h = 0.3$, and the mean bubble Reynolds number Re$_p \approx 136$. Spherical bubbles migrate toward the walls, but deformable bubbles distribute themselves nearly uniformly in the bulk of the flow. This is further elucidated by the tracks of the spherical (c) and deformable bubbles (d). Figures are adapted from \cite{dabiri2013transition}.}
	\label{fig:10_Triggie}
\end{figure*}

It is well known that bubbles or buoyant particles are inclined to a lift force when subjected to a mean shear in a flow. For spherical particles rising in a simple shear flow, the nature of these lift forces are by now well understood  \citep{loth2009equation,loth2008lift}. The case of bubbles, however, is quite different due to additional complexities arising from deformability, internal circulations, and surface contamination  \citep{loth2008quasi,van2007drag,magnaudet2000motion,legendre_magnaudet_1998,Dominique1997,takemura2002}. The focus in most situations is to assess the lateral forces which induce bubble migration towards or away from the wall, and this in turn can be expressed as a function of bubble properties~(deformability, size), and flow properties~(co-flowing channel, counter-flowing channel, or background turbulence level). 

Bubble laden wall-layers are a common observation, and have been extensively explored experimentally \citep{so2002laser,kitagawa2004experimental,van2006numerical,zhang2006numerical}. In the presence of turbulence, the two most common flow configurations are upward  \citep{nakoryakov1981local} and downward  \citep{kashinsky1999downward} turbulent channel flows. Although in single phase flow the two are identical, they differ greatly for two-phase situations, since the bubble buoyancy can now orient differently with respect to the mean shear near the channel walls. 
\cite{drew1993analytical} developed the earliest model to unveil the mechanisms involved.  Using an asymptotic analysis, the author was able to qualitatively reproduce the general trends for the velocity profiles and void fraction distributions. In comparison, DNS provide the ideal setting where the governing NS equations are solved numerically for both phases in such a way that all the length and time scales are fully resolved.   \cite{lu2006dns}  used DNS with front tracking to examine bubbly flows in a vertical channel. The results showed that for nearly spherical bubbles, the lift-induced lateral migration resulted in two regions: a nearly uniform velocity bulk flow region where the weight of the liquid/bubble mixture balances the imposed pressure gradient, and a wall layer that is free of bubbles for downflow and wall-rich for upflow \citep{triggvason2015direct}. The latter situation with bubble clustering near the walls is shown in {\bf Figure}~\ref{fig:10_Triggie}(a). While this result can be explained in the same spirit as the shear-induced lift of rigid spherical particles, strikingly, when the bubble is deformable the effect is reversed (see {\bf Figure}~\ref{fig:10_Triggie}(b)), and deformable bubbles distribute themselves in the bulk of the channel. The negligible lift for the deformable bubble can be attributed to the pliant nature of its interface, which prevents the buildup of a non-uniform surface pressure distribution. Thus, it is bubble deformability \citep{tomiyama2002transverse}, and not size,  which causes the sign change of the lift force in turbulent upflow channels and pipes. The flow modifications that the bubbles bring about in vertical channel flows will be reserved for a later section.



%

\section{TURBULENCE MODULATION BY BUBBLES}

In the preceding sections we reviewed a variety of dynamical regimes for bubbles and buoyant particles in turbulent flows. However, it is the collective behavior of these particle which often contributes to sizable effects in most engineering applications of particle-laden turbulence.  Next, we will discuss how some of the afore-described mechanisms play a crucial role in triggering the different kinds of flow modulations occurring in low to moderate volume fraction  ($\alpha < 5\%$) suspensions of bubble laden turbulence.


\subsection{Bubbles rising within  homogeneous turbulence}
\label{sec-wake_spectra}

\begin{figure*} [!tbp]
	\centering
	\includegraphics[width=0.8\linewidth]{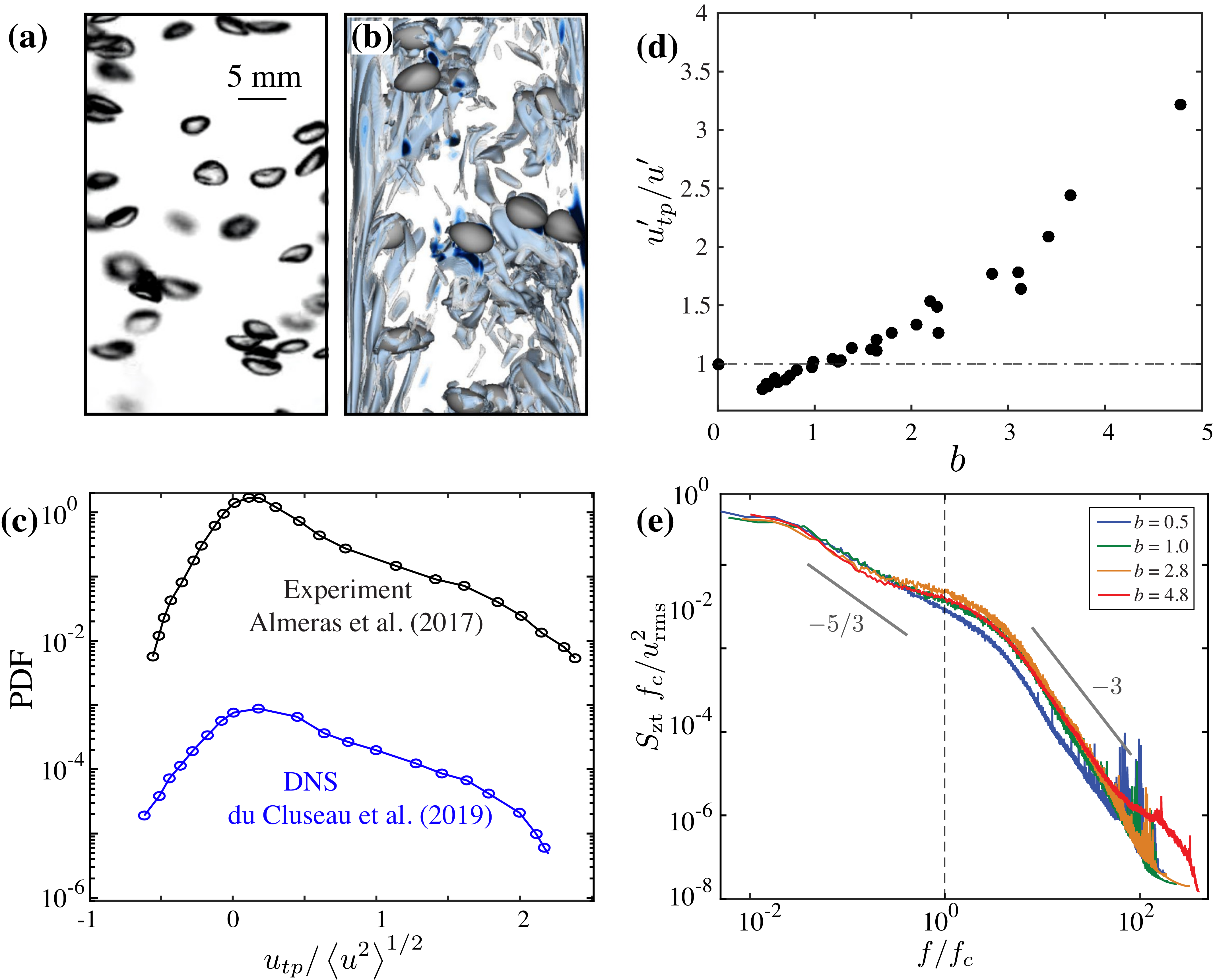}
	\caption{Flow modification in turbulent bubbly upflows. (a) Experimental images of 2-4 mm diameter deformable air bubbles in a vertical flow at Re$_\lambda = 262$. Snapshot adapted from the experiments reported in \cite{almeras2017experimental}. (b) Volume-rendering  of deformable bubbles in a turbulent flow, obtained from DNS with front tracking method. Figure  adapted from \cite{du2019analysis}. The flow structures are colored by isovalues of the so-called $\lambda_2$ criterion~ \citep{jeong1995identification}. (c) Probability density functions (vertically shifted for better visibility) of the axial velocity in the bulk of the flow from experimental~(black) and DNS~(blue) cases shown in (a) and (b), respectively. The PDF shapes are qualitatively similar in experiment and DNS. (d) Normalized vertical velocity fluctuations of the liquid phase for different values of the bubblance parameter $b$. (e) Normalized spectra of the bubble induced liquid fluctuations. Distinct ranges of scalings can be observed, i.e. $-5/3$ for $f/f_c  <1$ and $-3$ for $f/f_c >1$. $f_c$ is a characteristic cut-off frequency, which can be calculated using the bubble and flow properties. Figures (d) and (e) adapted from \cite{almeras2017experimental}.}
	\label{fig:11_turb_mod}
\end{figure*}

\subsubsection{Homogeneous isotropic turbulence}

{\color{black}Homogeneous and isotropic turbulence~(HIT) represents one of the most elementary forms of turbulence imaginable  \citep{batchelor1953theory}.  Notwithstanding how unrealistic this flow might seem from a practical viewpoint, studies of HIT have led to great breakthroughs in our understanding of real-world turbulent flows \citep{pop00}. The assumption of statistical homogeneity and isotropy (among others) has been central to many successful theories of turbulent flows.  It may be noted that all single phase flows, at high enough Reynolds numbers, will behave as HIT in the universal range and far from boundaries. For single phase turbulence, the $-5/3$ scaling of the energy spectrum (in the inertial range) is well known \citep{pop00}, {\color{black} wherein the energy flux flowing down to smaller length scales is nearly constant up to the dissipative scales.} In contrast, for a swarm of high Re$_p$ bubbles rising within an otherwise quiescent liquid, bubble induced turbulence (BIT) leads to a $-3$ scaling for the energy spectrum. This result, originally observed in the milestone work of \cite{lance1991turbulence}, has by now become well-established through detailed experiments, direct numerical simulations, and even large eddy simulations \citep{martinez2007measurement,roghair2011energy,riboux2013model}.} A further simplified approach was undertaken by \cite{Mazzitelli2009evolution}, using the PP equation of motion with an imposed back-reaction on the flow. This cumulative back reaction force ${\bf f_R} = \Sigma \big{(}\frac{D {\bf u}}{D t} - g \big{)}\mathcal{V}_p \ \delta({\bf x_f} - {\bf x_p})$, acting at the point ${\bf x_f}$ in the flow, did not generate the kind of liquid agitation commonly seen in BIT. Although the two-way coupling proved non-ideal for BIT, it demonstrated that the key ingredient crucial for the scaling (missing in the PP approach) was a model for the unsteady bubble wakes. For a detailed review of the liquid agitation induced by bubbles swarms~(BIT), readers are referred to \cite{risso2017agitation}.

Both single phase turbulence and bubbly swarms have been studied separately. But the situation where bubbles are injected into an already turbulent background flow, despite its relevance in industrial applications, has only recently begun to be understood. What determines the nature of liquid fluctuations and energy spectrum of such bubble laden turbulent flows? To allow comparisons across different levels of turbulence and bubble volume fractions, \cite{lance1991turbulence} and building on that \cite{rensen2005effect} introduced the so-called ``bubblance'' parameter $b$, which compares the intensity of BIT to the intensity of incident turbulence. This ratio of kinetic energies can be written as $b =  \alpha \bar{V_r}^2/u'^2$, where $\bar{V_r}$ is the mean rise velocity of the bubbles\footnote{Note that a prefactor of 1/2, based on the $C_M$ of a spherical bubble  \citep{van1998pseudo}, was used in the definitions of $b$ in \cite{rensen2005effect}}.  Recently, \cite{almeras2017experimental} conducted extensive experiments in such turbulent bubbly flows for a wide range of $b$. {\bf Figure}~\ref{fig:11_turb_mod}(a) shows a snapshot of 2-4 mm diameter bubbles rising in the bulk region of upward channel flow. The liquid velocity fluctuations in bubble laden turbulence was measured using a phase-sensitive hot-film anemometry  \citep{almeras2017experimental}, and shown to be positively skewed~(see {\bf Figure}~\ref{fig:11_turb_mod}(c)), as is also the case for BIT. Similarly skewed velocity PDF have been seen in DNS of bubbles in turbulent upflows \citep{du2019analysis} using the front tracking method~(see {\bf Figure}~\ref{fig:11_turb_mod}(b) and (c)). In the experiments, turbulence attenuation was observed at low values of $b$~($< 0.25$), while the liquid velocity fluctuations were augmented at larger $b$~(see {\bf Figure}~\ref{fig:11_turb_mod}(d)). {\color{black} It is important to stress that the liquid agitation produced by high Reynolds number bubbles is anisotropic \citep{risso2017agitation}, the effects of which are noticeable also in turbulent flows. For details about the anisotropy of liquid velocity fluctuations in bubble laden turbulence, the reader is referred to \cite{almeras2019mixing}. }


With the magnitude and distributions of liquid agitation revealed, the natural next question to ask is: How is the energy spectrum modified by the presence of bubbles in an incident turbulent flow? {\bf Figure}~\ref{fig:11_turb_mod}(e) shows the spectrum of liquid velocity fluctuations for $b$ in the range [0, 1.3]. With the addition of bubbles, the higher frequencies of the inertial sub-range of single phase turbulence are substituted by a $-3$ scaling of BIT. However, the -5/3 scaling appears to be preserved for the lower frequencies for all values of $b$ tested. The characteristic cut-off frequency $f_c $ imposed by the bubble swarm may then be calculated as $ \bar{V_r}/\lambda_c$, where $\lambda_c = d_p/C_{d0}$  where $C_{d0}$ is the drag coefficient of an isolated bubble in still fluid  \citep{riboux2010experimental}. The above results only partially resolve the complexity of the problem. One may note that the spectrum modification by 2-4 mm diameter bubbles (reported above) is at variance with some of the observations in   \cite{risso2017agitation}. The reason for this lies in the differences in the operating conditions, i.e. the size and  Reynolds number of the bubbles. With a change in bubble diameter, the cut-off length scale $\lambda_c$ is different. This leaves room for much variability in the frequency range of BIT. These issues can be resolved through careful studies where bubble size and bubble Reynolds number are controlled independently. Such studies will however be challenging.

\subsubsection{Homogeneous shear turbulence}

Homogeneous shear turbulence can be considered one of the simplest turbulent flows where the flow relaxes the condition of statistical isotropy, but maintains homogeneity over all spatial scales  \citep{pumir1996turbulence,champagne1970experiments}. Explorations of bubble-laden homogeneous shear turbulence have been largely on the simulations side  \citep{gualtieri2015transport}. As a general rule, bubbles enhance the dissipation rate of turbulent kinetic energy  \citep{rosti2018droplets}; however, its production rate can be either enhanced or diminished depending on the flow parameters  \citep{kawamura2006direct}. While the TKE production rate increases with the turbulent Reynolds number, it was found to decrease with the shear Reynolds number. In addition, the effect of bubble deformability was to enhance the TKE production rates.  Although simplified models have been proposed to explain these, the absence of direct experimental evidence leaves much debate on the validity of these mechanisms in causing turbulence modulation.

\subsection{Bubble laden channel flows}

\subsubsection{Horizontal channels and boundary layers}

The injection of bubbles, small or large, in turbulent boundary layers and horizontal channel flows has long been known to modulate the flow, with several experimental explorations in the past few decades  \citep{madavan1984reduction,madavan1985measurements,gutierrez2008turbulence,sanders2006bubble}. Recent numerical works include those due to  \cite{xu2002numerical} and   \cite{ferrante2004physical} employing Eulerian-Lagrangian models, and those due to \cite{lu2005effect} using DNS to reveal how bubbles can modify the near-wall vortical structures causing friction drag reduction even with relatively low volume fractions. Reviews on drag reduction by bubbles in turbulent boundary layers can be found in    \cite{murai2014frictional},   \cite{ceccio2010friction}, and more recently in  \cite{rawat2019drag}.

\begin{figure*} [!tbp]
	\centering
	\includegraphics[width=0.99\linewidth]{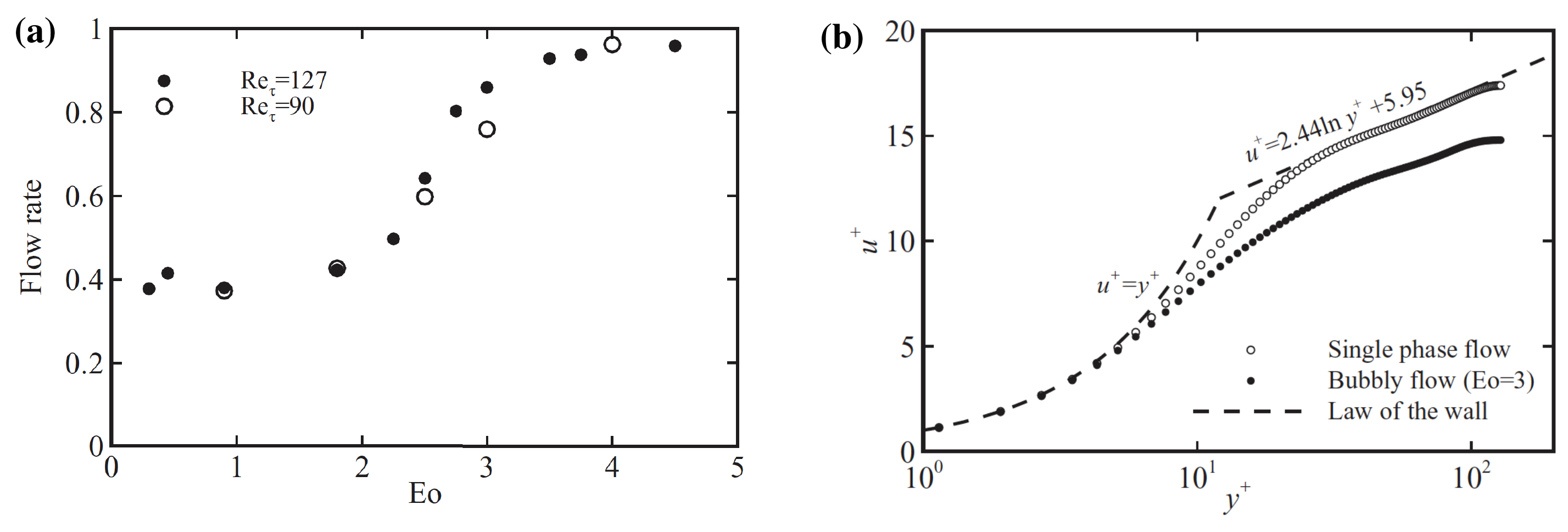}
	\caption{Flow modification in turbulent bubbly upward channel flows for  3\% volume fraction of bubbles with a size ratio $d_p/h = 0.3$, where $h$ is the channel half width. (a) Flow rate as a function of bubble deformablity, quantified here using E\"otv\"os number Eo (or equivalently the Bond number Bo), for two different values of friction Reynolds number Re$_{\tau}$. The friction Reynolds number effect is negligible; the phenomenon is governed primarily by the bubble deformability. Bubble Reynolds number Re$_p $ lies in the range [120, 200]. (b) Velocity profile near the wall for the Eo = 3~(Re$_\tau = 127$) case in (a), as compared to the single phase case. Figures adapted from \cite{dabiri2013transition}.}
	\label{fig:12_verti_turb_mod}
\end{figure*}

\subsubsection{Vertical channels}

Turbulence modulation in bubbly vertical channel flows has historically been the subject of numerous investigations, both experimental  \citep{serizawa1975turbulence} and analytical  \citep{antal1991analysis}. As earlier introduced in section \ref{lift_lateral}, with high resolution DNS becoming increasingly  feasible, simulations employing the front tracking method  \citep{unverdi1992front} have led to significant breakthroughs in our understanding of these problems  \citep{lu2008effect,du2019analysis,tryggvason2011direct,lu2006numerical,dabiri2016scaling}. Here we will highlight the main effects caused by bubbles on the underlying flow field.  For a bubble laden channel flow, the relevant dimensionless parameters are the channel Reynolds number Re$_h$, the bubble Galileo number Ga, and the bubble E\"otv\"os number Eo. The bubble Reynolds number, which is an output parameter Re$_p$, can be expressed as a function of Ga, Eo, and, to a weaker extent, the volume fraction $\alpha$.  For upward channels flows, the most prominent effect is a reduction in the net flow rate. The flow rate reduction is most extreme for spherical bubbles, since they occupy the near wall regions. Highly deformable bubbles, owing to their nearly zero (or slightly negative) lift force, remain in the bulk of the flow, thus having negligible effects on the volumetric flow rate~(see {\bf Figure}~\ref{fig:12_verti_turb_mod}(a)). The physical mechanism behind  the flow rate reduction is a sudden rise in the near-wall viscous dissipation when the spherical bubbles enter the viscous sublayer. As dictated by their relative distributions, the near-wall liquid velocity fluctuations are enhanced for spherical bubbles, but the same occurs in the bulk for deformable bubbles. In addition the turbulent velocity fluctuations in the bubble rich regions are enhanced, a result which directly follows from the knowledge of BIT  \citep{risso2017agitation}. The velocity profile near the wall is retarded in the presence of bubbles~(see {\bf Figure}~\ref{fig:12_verti_turb_mod}(b)), which is consistent with the observed flow rate reductions. In the less commonly explored downflow configuration with bubbles  \citep{lu2006dns}, a flow rate reduction is accompanied by a suppression of turbulence in the near wall bubble-free layer, and turbulence augmentation in the bubble-rich bulk flow (see also section \ref{lift_lateral}). Thus, while in many engineering applications it might seem beneficial to reduce the bubble size and maximize the interfacial area, such efforts should not be made at the expense of a flow rate reduction due to viscous losses of the bubbles accumulating near the walls.

Lastly, we note that the above discussions pertain to clean bubbles. In reality, the void fraction profiles in turbulent bubbly upflow experiments  \citep{serizawa1975turbulence} are possibly not as sharply peaked as seen in the simulations~(DNS). When the gas-liquid interfaces are contaminated (with dirt or surfactants), a slight reduction of the lift force and lateral migration tendency can be expected  \citep{lu2017effect,clift1978bubbles,takagi2011surfactant}. {\color{black}Thus, even at low volume fractions of the dispersed phase, bubble dynamics can show fundamental differences to that of particles due to the presence of an internal circulation. On the other hand, dense bubbly flows ($\alpha \geq 5\%$) need to be treated as a markedly different subject; for a review of experimental work on bubbles in vertical pipe/channel flows, see \cite{guet2006fluid}.} Reviews discussing flow regimes, operating parameters and design parameters of industrial bubble columns can be found in  \cite{besagni2016comprehensive}.


\begin{figure*} [!tbp]
	\centering
	\includegraphics[width=0.9\linewidth]{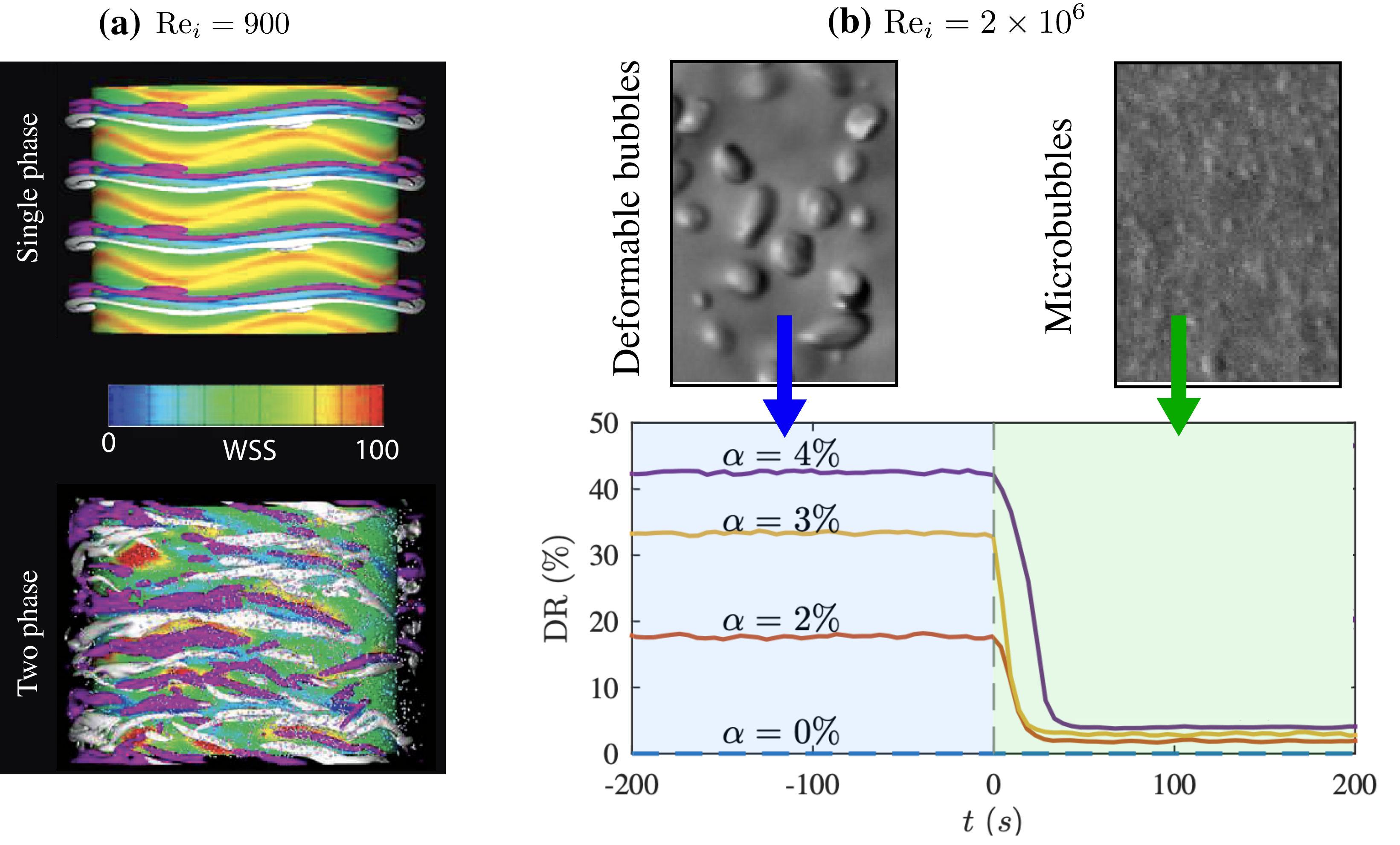}
	\caption{Bubbly drag reduction~(DR) in turbulent Taylor-Couette flow. (a) Direct numerical simulations of Taylor-Couette~(TC) turbulence at inner cylinder Reynolds number Re$_i$ = 900 for single-phase flow~(top)  and two-phase flow laden with point-like bubbles (bottom). Wall shear stress (colorbar) is reduced in the two phase flow with small bubbles, leading to about 20\% DR. Figures adapted from \cite{sugiyama2008microubbly}. (b) DR\% as function of time at a high Re$_i = 2 \times 10^6$. Nearly all DR is lost after
		injection of a surfactant~(Triton X-100), which reduces the two phase flow from a suspension of large deformable bubbles~(upper left inset) to microbubbles~(upper right inset). Data and figures from \cite{verschoof2016bubble}.}
	\label{fig:13_TC_DR}
\end{figure*}

\subsection{Bubbly Taylor-Couette turbulence}

Taylor-Couette (TC) flow, the flow between two coaxial co- or counter-rotating cylinders, is one of the paradigmatic systems of fluid physics \citep{grossman2016high}.  Unlike the turbulent flows discussed above, the TC flow is a closed system with exactly derivable balances between driving and dissipation. Owing to these particular benefits, turbulent TC flow experiments have been widely used by researchers as a model system to study bubble induced flow modulations, bubble-vortex interactions and bubbly drag reduction  \citep{murai2005bubble,murai2008frictional,berg2005,van2013importance,chouippe2014,fokoua2015,verschoof2016bubble}. A remarkable effect of introducing bubbles in turbulent TC flow is that with only a small percentage ($\alpha \sim \mathcal{O}(1\%)$) of bubbles, significant turbulence modulation can be achieved. This usually manifests in major drag reduction, with various mechanisms contributing to it, see also section 9 of \cite{lohse2018bubble}.

\subsubsection{Drag reduction in the buoyancy dominant regime}
We first discuss the studies on drag reduction by microbubbles in turbulent TC flow. 
To this end, \cite{sugiyama2008microubbly} and \cite{spandan2016drag} conducted DNS employing an Euler-Lagrangian~(PP) two-way coupled approach in a regime where  the inner cylinder Reynolds number Re$_\text{ic} \equiv \omega_\text{ic} r_\text{ic} (r_\text{oc} - r_\text{ic})/\nu$ was in the range [600, 8000]. Here $\omega_\text{ic}$, $r_\text{ic}$, and $r_{\text{oc}}$ are the inner cylinder angular velocity, inner cylinder radius, and outer cylinder radius, respectively. The drag reduction was defined as 
\begin{equation}
\text{DR}(\%) = \frac{\left < C_f \right >_s - \left < C_f \right >_\text{tp}}{\left < C_f \right >_s} \times 100,
\label{eq_cf_spandan}
\end{equation}
where $C_f = ((1-r_\text{ic}/r_\text{oc})^2/\pi)G/\text{Re}_\text{ic}^2$ is the friction factor, $G = { \tau\over 2\pi \ell_c \rho_l \nu^2 } $ is the dimensionless torque, with $\tau$ the torque that is necessary to keep the inner cylinder of length $\ell_c$ rotating at constant angular velocity \citep{zhu16}.
The subscripts `$s$' and `$\text{tp}$' denote single phase and two phase, respectively. The simulations of \cite{sugiyama2008microubbly} and \cite{spandan2016drag} showed that the buoyant motions of microbubbles can disrupt the coherent vortices (``Taylor rolls"), resulting in a reduction of drag (up to 20 \%) on the inner cylinder surface~(see {\bf Figure}~\ref{fig:13_TC_DR}(a)). \cite{spandan2016drag} varied the Froude number Fr$_\text{ic} = \omega_{\text{ic}} \sqrt{r_\text{ic}/g}$, representing the ratio of centripetal force strength over buoyancy, in the range  [0.16, 2.56]. Keeping the values of Re$_\text{ic}$, $\alpha$ and $d_p$ fixed, drag reduction was found to be significant at low Froude number (Fr$_\text{ic} \lesssim 1$) and negligible at high Fr$_\text{ic} \gtrsim 1$. In this regime, the drag reduction becomes smaller with increasing Reynolds number. For a more generic quantification of DR in this low Reynolds number regime, one might incorporate the bubble Stokes number and volume fraction as well, in which case the DR\% can likely be expressed as a function of St/Fr and $\alpha$ (assuming the bubbles are non-inertial). The hope is to obtain an overarching dimensionless parameter that can explain the degree of drag reduction in the so-called wavy vortex regime \citep{marcus1984simulation,fardin2014hydrogen,and86} of TC turbulence.

\subsubsection{Drag reduction in highly turbulent regime}
In the large Re$_\text{ic}$ regime, the stable coherent structure of the vortices in TC turbulence gets lost. Consequently, the effect of the bubbles on the friction drag diminishes, as mentioned above. However, bubbles can still be used to reduce drag in the highly turbulent TC regime. In this high Reynolds number regime, however, the deformability of the bubble (i.e., large Weber number $\text{We} > 1$) is crucial for DR \citep{berg2005,van2013importance}. In this large Reynolds number regime, drag reduction increases with increasing Re$_\text{ic}$. 
A direct experimental demonstration of the effects of bubble deformability on turbulent TC drag reduction can be found in \cite{verschoof2016bubble}. These authors dynamically changed the drag by adding a minute amount of surfactant~(Triton X-100) to a highly turbulent TC flow~(Re$_\text{ic}$ up to 2 $\times 10^6$) laden with deformable bubbles~(see Figure~\ref{fig:13_TC_DR}(b)). In the original state with only a 4\% volume fraction of deformable bubbles, the DR\% was over 40\%~(left half of the figure). The addition of surfactant initiated a remarkable turn of events (breakup, coalescence prevention, etc) that caused the large deformable bubbles to be fully substituted by tiny microbbubles~(right half of the figure), thereby reducing the DR to just 4\%, which corresponds to the trivial effect of the bubbles on the density and viscosity of the liquid.

Recent work by \cite{spandan2018physical} have used DNS to investigate the physical mechanisms of drag reduction in the turbulent regime~(up to Re$_\text{ic} = 2 \times 10^4$). They connected the increase in drag reduction to a decrease in the dissipation in the wake of highly deformed bubbles near the inner cylinder. This touches the familiar territory of polymer drag reduction  \citep{benzi2018polymers,procaccia2008RMP,white2008}, and indicates interesting similarities in DR mechanisms where elastic properties of the dispersed phase are being exploited.  

Yet another important issue when studying bubbly drag reduction in TC flow is the effect of centripetal force on the bubble distribution in the flow. 
If the TC flow is in a fully laminar state~(and Fr$_\text{ic}  \gg$ 1), all the bubbles should be pushed against the inner wall due to the radial pressure gradient induced by the centrifugally driven flow. However, when the system is a highly turbulent state, the bubbles experience liquid velocity fluctuations and pressure fluctuations, which are enough to diffuse them towards the bulk region of the flow \citep{van2013importance}. The resulting bubble distribution in the gap between the cylinders will depend on the competing effects of the turbulent pressure fluctuations induced acceleration $a_\text{pf}$ and the centripetal acceleration $a_\text{{c}}(r)$. This can be defined as a so-called\footnote{Note that this Froude number should not be mistaken with the more widespread definition of Froude number where gravity force appears in the denominator. Here, the body force in the denominator is centrifugal.} centripetal Froude number
\begin{equation}
\text{Fr}_c(r)=\frac{a_\text{pf}}{a_{c}(r)}=\frac{{u'}^2/d_{p}}{U_{\theta}^2/r},
\label{eq:Fr_cent}
\end{equation}
with $U_{\theta}$ the mean azimuthal liquid velocity and $r$ the radial position in the TC setup under consideration. \cite{van2013importance} estimated that Fr$_c \approx 1.6$ at Re$_\text{ic}$ $\sim$ $5 \times 10^5$ and  Fr$_c \approx 3.4$ at Re$_\text{ic} = 1 \times 10^6$. The lower Re$_\text{ic}$ implies a lower Fr$_c$  and thus the effective centripetal force on the
bubbles is higher and, hence, the bubble accumulation is stronger near the inner cylinder wall at the expense of a lower concentration in the bulk. This reasoning is also consistent with their direct experimental observations.

Thus to summarize, drag reduction in bubbly Taylor-Couette turbulence is a function of several parameters. While at moderate Re$_\text{ic}$ the buoyancy-induced drift of the microbubbles are sufficient, in the highly turbulent regime, buoyancy, deformability and centripetal effects are all  crucial ingredients to DR. In  light of the close analogy that exists between TC flows and pipe flows \citep{eck07a}, the results obtained in turbulent TC flows are of value to drag reduction research. Yet, whether and how the principles of turbulent two-phase TC flow can be extended to pipelines and naval applications needs to be astutely examined.

\subsection{Bubbles in turbulent convection }

\begin{figure*} [!tbp]
	\centering
	\includegraphics[width=0.99\linewidth]{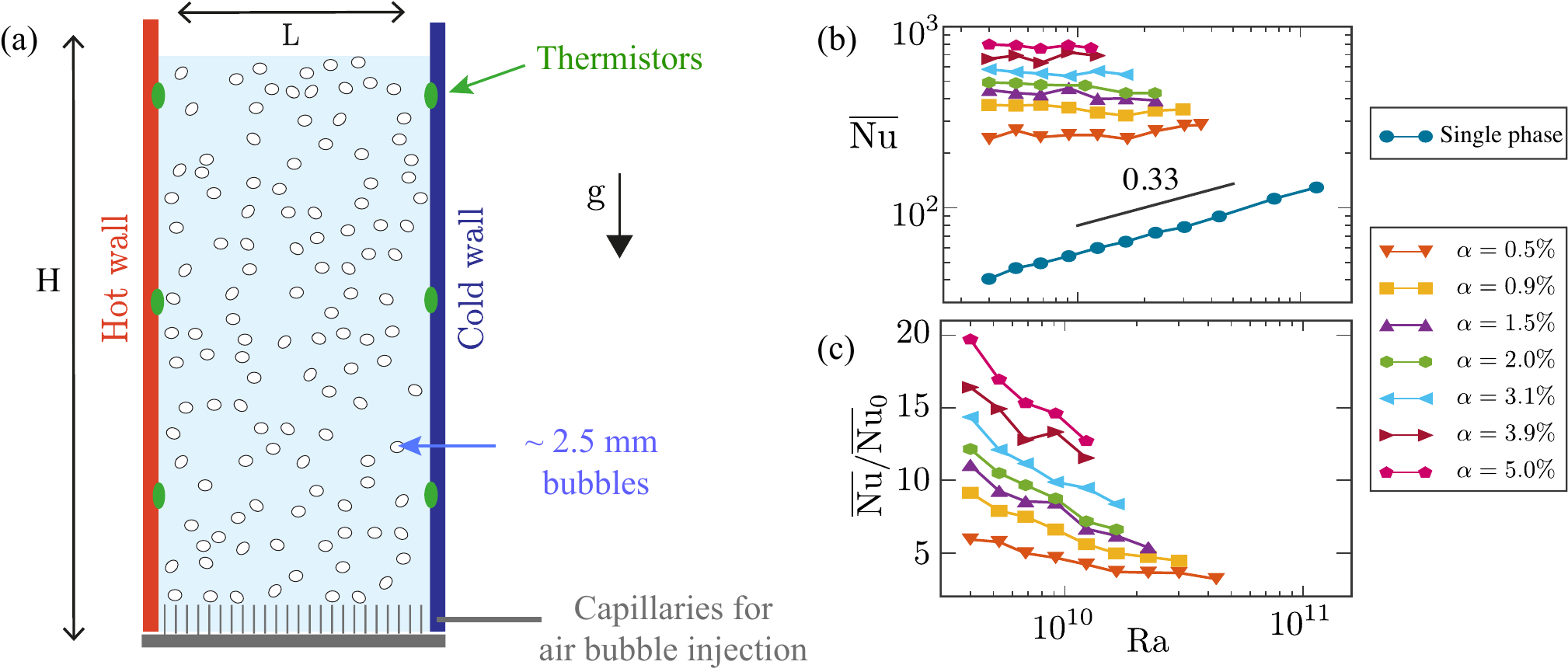}
	\caption{(a) A sketch of a turbulent vertical convection setup (experiment) with bubble injection. Rectangular bubbly column heated from one sidewall and cooled from the other side, where the height $H$ = 600 mm, and length between heated and cooled side walls $ L$ = 230 mm. Bubbles of diameter of about 2.5~mm were injected into the system through 180 capillaries (inner diameter 0.2 mm) placed at the bottom of the apparatus. (b) Nusselt number $\overline{\text{Nu}}$ in bubble laden vertical convection versus  Rayleigh number $\text{Ra}_H$  for various gas volume fractions $\alpha$. Blue circles shown for comparison correspond to the single phase turbulent vertical convection case~($\alpha = 0\%$). The size of the symbols corresponds to the typical error-bar in the data. (c) Heat transfer enhancement due to bubble injection $\overline{\text{Nu}}/\overline{\text{Nu}}_0$ versus Ra$_H$. Here, $\overline{\text{Nu}}_0$  is the Nusselt number of the single-phase case. Figure adapted from \cite{gvozdic2018experimental}.}
	\label{fig:14_VN_HT}
\end{figure*}

In heat transfer systems, the motion of bubbles is known to be able to efficiently induce mixing of warm and cold parcels of liquid. For many industrial applications, injecting bubbles in the flow can lead to a 100 times enhancement in the heat transfer coefficient as compared to its  single-phase counterpart \citep{deckwer1980mechanism}.  Therefore, the effect of bubbles and light particles on heat transfer has been subject of several experimental and numerical investigations. One approach for enhancing heat transport is to create vapor bubbles \citep{prosperetti2017vapor} or biphasic \citep{wang2019self} and thermally expandable particles \citep{alards2019statistical}.
Some of these have resulted in impressive heat transport enhancements as compared to the single phase case \citep{zhong2009enhanced,lakkaraju2013heat,wang2019self}. However, in the present review, we restrict our attention to gas phase bubbles. 

Early studies in forced convection systems with bubble injection \citep{sekoguch1980forced, sato1981momentum, sato1981momentum2} showed that the bubbles modify the temperature profile of the system, and the gas volume fraction close to the heated wall is important for heat transfer enhancement, i.e. higher void fractions close to the heated wall lead to an enhanced heat transfer. Using DNS with front tracking, \cite{dabiri2015heat} recently studied the effect of bubbles on the heat transfer rate in a flow between two parallel walls under a constant heat flux condition. They found that the bubbles stir up the viscous layer and reduce the size of the conduction region near the wall, resulting in an improved heat transfer efficiency, i.e. a 3\% volume fraction of bubbles can increase the Nusselt number by 60\%.

Studies on the effects of bubble injection on heat transfer in natural convection systems were mostly conducted with injection of micro-bubbles \citep{kitagawa2008heat,kitagawa2009effects} and sub-millimeter-bubbles \citep{kitagawa2013natural} close to the heated wall. 
\cite{kitagawa2013natural} investigate the effects of microbubble injection on natural convection heat transfer from a vertical heated plate in water. They found microbubble injection significantly increases the heat transfer coefficient in both the laminar and transition regimes. The enhancement ratio of the heat transfer coefficient due to bubble injection is 1.6-2.0 in the laminar regime and 1.5-2.0 in the transition regime. The physical reason for heat transfer enhancement in the laminar regime is due to effective mixing, whereas the physical reason for the heat transfer enhancement in the transition region is because microbubble injection accelerates the transition to turbulence \citep{kitagawa2013natural}.
\cite{deen2013direct} studied wall-to-liquid heat transfer
in dispersed gas liquid two-phase flow using DNS, and found that a few high Reynolds number bubbles rising in quiescent liquid can considerably increase the local heat transfer between the liquid and a hot wall. 

Recently, \cite{gvozdic2018experimental} studied the effect of deformable bubbles (with diameter of 2-3 mm) on heat transfer in a vertical natural convection setup, which was heated from one side and cooled from the other side (as shown in {\bf Figure}~\ref{fig:14_VN_HT}(a)).
The air bubbles were injected into the system using 180 capillaries (inner diameter 0.21 mm) uniformly distributed over the bottom of the nature convection system.  The gas volume fraction $\alpha$ varied from 0.5\% to 5\%, and the Rayleigh number ranged from $4.0 \times 10^9$ to utmost $3.6 \times 10^{10}$. Here, Rayleigh number is defined as $\text{Ra}_H =  \frac{g  \gamma  (\overline{T_h}-\overline{T_c})  H^3}{\nu   \kappa_c}$, and the Nusselt number as $\overline{\text{Nu}} = \frac{Q_c / A}{K \ (\overline{T_h} - \overline{T_c})/L}$, 
where $\gamma$ is the thermal expansion coefficient, $\overline{T_h}$ and $\overline{T_c}$ are the mean temperatures of the hot and cold walls, respectively, $L$ is the length of the setup, $A$ is the surface of the sidewall,  $\kappa_c$ the thermal diffusivity, $K$ the thermal conductivity of water, and $Q_c$ is the measured power supplied to the heaters. 
For the entire range of $\alpha$ and $\text{Ra}_H$, adding bubbles dramatically increased the heat transport efficiency, as the Nusselt number is about an order of magnitude higher as compared to single-phase flow case ({\bf Figure}~\ref{fig:14_VN_HT}(b)). 
In order to more clearly quantify the heat transport enhancement due to bubble injection, {\bf Figure}~\ref{fig:14_VN_HT}(c) shows the ratio of the Nusselt number for two phase bubbly flow, $\overline{\text{Nu}}$, to that of the single-phase case, $\overline{\text{Nu}_{0}}$, as a function of $\text{Ra}_H$ at different $\alpha$. 
It is shown that heat transfer was enhanced up to 20 times thanks to the bubble injection, and that the heat transfer enhancement increased with increasing the gas volume fraction $\alpha$ and decreasing Rayleigh number $\text{Ra}_H$. 
Note that the decreasing trend of $\overline{\text{Nu}}/\overline{\text{Nu}_0}$ with $\text{Ra}_H$ occurred because the single-phase Nusselt number $\overline{\text{Nu}_0}$ increased with $\text{Ra}_H$ whereas the two phase Nusselt number did not change with $\text{Ra}_H$. It was found that the Nusselt number $\overline{\text{Nu}}$ was nearly independent of $\text{Ra}_H$ and depended solely on the gas volume fraction~$\alpha$ with the scaling of $\overline{\text{Nu}} \propto \alpha^{0.45}$, which is suggestive of a diffusive transport mechanism as found in the case of the mixing of a passive tracer in a homogeneous bubbly flow for a low gas volume fraction (\cite{almeras2015mixing,almeras2019mixing}). Thus, bubble-induced mixing dominates the efficiency of the heat transfer in the moderate Ra$_H$ bubbly nature convection systems.



\begin{summary}[SUMMARY POINTS]
	\begin{enumerate}
		\item The past decade has witnessed tremendous progress in our understanding of  buoyant particle  and bubble laden turbulent flows. The addition of buoyant particles to turbulent flows can modify key aspects of single phase turbulence, such as spectra or drag. This offers opportunity to employ bubbles or light particles to tailor turbulent flows to our benefit.

		\item As in many areas of modern fluid dynamics, fully resolved direct numerical simulations offer great potential, capable of explaining many intricate phenomena of two-phase turbulence. At the same time, a reduced treatment employing the Euler-Lagrangian approach prove sufficient in a remarkably large number of situations. The basic point particle formulation have been extended to include rotation through the Kelvin-Kirchhoff equations, which have significant predictive capabilities for buoyant particle and bubble dynamics in flows.
		
		\item Deformable bubble dynamics in turbulence is found to be governed by three fairly independent mechanisms, which are as follows: The average bubble shape is mainly controlled by the relative motion between the bubble and the surrounding fluid; the bubble velocity and orientation are a result of its own wake instability; and the effect of turbulence reflects through random deformations of the bubble interface, which under extreme situations can cause bubble breakup.
		
		\item Buoyancy brings about a multitude of modifications to particle dynamics in turbulence. For small bubbles  and particles the crossing trajectories effect leads to augmented particle accelerations, while in finite sized and finite particle Reynolds number cases, the wake induced accelerations add to the turbulent forcing. When the particle Reynolds number is increased further, the buoyant particle's rotation further aids in the development of vigorous accelerated motions.

		\item Adding bubbles to turbulent flow is not synonymous with drag reduction. While bubbly drag reduction is possible in horizontal channel flows, boundary layers and Taylor-Couette flows, in vertical channel flows (both upflows and downflows) the effective drag is enhanced.   The reason is that in vertical channels the bubbles increase the energy dissipation rate, while in Taylor-Couette turbulence and other flows they result in  suppression of the dissipation.

		\item \textcolor{black}{Air bubbles added to (open) turbulent convection systems  dramatically enhance the heat transfer, thanks to their induced liquid agitation and mixing. }
	\end{enumerate}
\end{summary}

\begin{issues}[FUTURE DIRECTIONS]
	\begin{enumerate}

		\item Rigid buoyant anisotropic particles, including ellipsoidal, chiral and vaned particles, can add significant amounts of energy  to turbulent flows. Their coupled translational-rotational dynamics is crucial to the liquid agitation.
		
		\item Varying rotational inertia and/or center of mass location of buoyant particles presents exciting opportunities for turbulence modulation.
		
		\item Whether the collective wake instabilities of rising bubbles and buoyant particles persist in intense turbulent environments or not is an open question.
		
		\item The issue of energy spectra in bubble laden turbulence is only partially resolved. The effects of the bubble size as compared to the Kolmogorov scale is not clear. Similarly, the spectrum modification by the wakes of low to moderate Reynolds number bubbles remains to be elucidated. Furthermore, the behavior of bubble laden turbulent flows in  the limit of very large Reynolds number turbulence needs to be studied.
		
		\item Tumbling buoyant particles can be engineered for turbulent downflow channels to affect the near wall turbulence, with a potential for heat transfer enhancements. Bubbles cannot be used to achieve this.

		\item The response of bubbles to homogeneous shear turbulence remains to be experimentally explored. This requires the design of dedicated experimental setups.  The relative alignment between buoyancy and shear, in combination with deformability is expected to induce symmetry breaking, and rich variability in bubble dynamics can be expected.

	\end{enumerate}
\end{issues}

\section*{DISCLOSURE STATEMENT}
The authors are not aware of any affiliations, memberships, funding, or financial holdings that might be perceived as affecting the objectivity of this review. 

\section*{ACKNOWLEDGMENTS}
We thank all coworkers for their contributions and for the many stimulating discussions over the years. The authors acknowledge support from the Natural Science Foundation of China under grant nos 11988102, 91852202, 11861131005 and 11672156, the Max Planck Center Twente for Complex Fluid Dynamics, NWO, and European Research Council~(ERC) via an Advanced Grant for financial support over the years.
%

\end{document}